\documentclass[oneside,11pt]{article}

\usepackage[utf8]{inputenc}
\usepackage[T1]{fontenc}

\usepackage{amsthm}
\usepackage{amsmath}
\usepackage{amsfonts}
\usepackage{amssymb}

\usepackage{graphicx} 
\usepackage{float}
\usepackage{booktabs}
\usepackage{multirow}
\usepackage{rotating}
\usepackage{array}
\usepackage{xcolor}
\usepackage{subfig}
\usepackage[ruled]{algorithm2e}

\usepackage{breakcites}

\usepackage{nowidow}

\usepackage{enumitem}

\usepackage[top=1.5cm,bottom=2cm,right=2.5cm,left=2.5cm]{geometry}
\usepackage{titling}

\usepackage{natbib}

\usepackage{ifplatform}

\usepackage{textcomp}

\ifwindows
\fi

\ifwindows
	\usepackage{bbm}
\else
	\iflinux
		\usepackage{bbm}
	\else
		\usepackage{bbm}
	\fi
\fi

\DeclareFontFamily{OT1}{pzc}{}
\DeclareFontShape{OT1}{pzc}{m}{it}{<-> s * [1.10] pzcmi7t}{}
\DeclareMathAlphabet{\mathpzc}{OT1}{pzc}{m}{it}

\usepackage[pdftex,final,bookmarksnumbered,bookmarksopen=false,breaklinks,colorlinks]{hyperref}
\hypersetup{
	final,
	bookmarksnumbered=true,
	bookmarksopen=true,
	bookmarksopenlevel=0,
	unicode=false,          
	pdftoolbar=true,        
	pdfmenubar=true,        
	pdffitwindow=false,     
	pdftitle={A goodness-of-fit test for the functional linear model with functional response},    
	pdfdisplaydoctitle=true, 
	pdftoolbar=true,
	pdfmenubar=true,
	pdflang={English},
	pdfauthor={Eduardo Garcia-Portugues, Javier Alvarez-Liebana, Gonzalo Alvarez-Perez, Wenceslao Gonzalez-Manteiga},     
	pdfsubject={arXiv paper},   
	pdfcreator={Eduardo Garcia-Portugues},   
	pdfproducer={Eduardo Garcia-Portugues}, 
	pdfkeywords={Bootstrap}{Cram\'er--von Mises statistic}{Functional data}{Regularization}, 
	pdfnewwindow=true,      
	breaklinks=true,
	hidelinks,       
	linkcolor=black,          
	citecolor=black,        
	filecolor=magenta,      
	urlcolor=cyan           
}

\newtheorem{remark}{Remark}

\newtheorem{algo}{Algorithm}

\newtheorem{lemma}{Lemma}

\allowdisplaybreaks

\newif\ifmain
\maintrue
\ifmain

\else

\fi
\newif\ifsupplement
\supplementtrue
\newif\iffigstabs
\figstabstrue

\begin{document}

\ifmain

\title{A goodness-of-fit test for the functional linear model\\ with functional response}
\setlength{\droptitle}{-1cm}
\predate{}%
\postdate{}%

\date{}

\author{Eduardo Garc\'ia-Portugu\'es$^{1,2}$, Javier \'Alvarez-Li\'ebana$^{3}$,\\
	Gonzalo \'Alvarez-P\'erez$^{4,6}$, and Wenceslao Gonz\'alez-Manteiga$^{5}$}

\footnotetext[1]{Department of Statistics, Carlos III University of Madrid (Spain).}
\footnotetext[2]{UC3M-Santander Big Data Institute, Carlos III University of Madrid (Spain).}
\footnotetext[3]{Department of Statistics and Operations Research and Mathematics Didactics, University of Oviedo (Spain).}
\footnotetext[4]{Department of Physics, University of Oviedo (Spain).}
\footnotetext[5]{Department of Statistics, Mathematical Analysis and Optimization, University of Santiago de Compostela (Spain).}
\footnotetext[6]{Corresponding author. e-mail: \href{mailto:gonzaloalvarez@uniovi.es}{gonzaloalvarez@uniovi.es}.}

\maketitle


\begin{abstract}
The Functional Linear Model with Functional Response (FLMFR) is one of the most fundamental models to assess the relation between two functional random variables. In this paper, we propose a novel goodness-of-fit test for the FLMFR against a general, unspecified, alternative. The test statistic is formulated in terms of a Cram\'er--von Mises norm over a doubly-projected empirical process which, using geometrical arguments, yields an easy-to-compute weighted quadratic norm. A resampling procedure calibrates the test through a wild bootstrap on the residuals and the use of convenient computational procedures. As a sideways contribution, and since the statistic requires a reliable estimator of the FLMFR, we discuss and compare several regularized estimators, providing a new one specifically convenient for our test. The finite sample behavior of the test is illustrated via a simulation study. Also, the new proposal is compared with previous significance tests. Two novel real datasets illustrate the application of the new test.
\end{abstract}
\begin{flushleft}
\end{flushleft}

\section{Introduction}
\label{sec:1}

The increasing availability of data for continuous processes has boosted the field of Functional Data Analysis (FDA) in the last decades as a powerful tool to take advantage of the complexity and rich structure of this kind of data, difficult to manage for many traditional statistical techniques given their intrinsically infinite dimensionality. Some of the main monographs in FDA are \cite{RamsaySilverman05}, \cite{FerratyVieu06}, \cite{HorvathKokoszka12}, and \linebreak\cite{HsingEubank15}.\\

Regression models with functional covariates and/or responses emerged as natural generalizations of multivariate ones. A specific instance arises when assessing the relation between two functional random variables $\mathcal{X}$ and $\mathcal{Y}$ via a general regression model $\mathcal{Y}=m(\mathcal{X})+\mathcal{E}$, where $\mathcal{E}$ is a functional random error. The main difference with the multivariate case is that here $m$ is an operator between function spaces, typically of a Hilbertian nature, therefore generalizing the usual Euclidean-Euclidean regression mapping. Nonparametric estimation of $m$ was addressed by \cite{Ferratyetal11} and \cite{lian2011}, who investigated the rates of convergence of kernel and $k$-nearest neighbors regression estimates, respectively. Moreover, \cite{Ferratyetal12} studied the nonparametric estimation of $m$ by considering data-driven bases and consistent bootstrap approaches.\\

However, much of the existing regression literature is concerned with (infinite-dimensional) parametric modeling, where the operator $m$ is assumed to belong to a given parametric family. As an early precedent, the simplest and best-known paradigm is the Functional Linear Model with Scalar Response (FLMSR), $Y=m_{\rho}(\mathcal{X}) + \varepsilon$, where $\varepsilon$ is a real-valued error and $m_{\rho}$ is a linear functional depending on a function $\rho$. Within the FLMSR, the so-called Functional Principal Components Regression (FPCR) was introduced by \cite{Cardotetal99} as a parsimonious estimation approach. \cite{Crambesetal09} proposed a smoothing splines estimator, whereas \cite{AguileraAguilera13} formulated penalized FPCR estimation techniques based on B-splines. Alternatively, functional partial least squares regression was proposed in \cite{PredaSaporta05}. Some authors have also studied the relation of a functional response and a scalar regressor, see, e.g., \linebreak\cite{Chiouetal03}.\\

In contrast, the Functional Linear Model with Functional Response (FLMFR), $\mathcal{Y}=m(\mathcal{X})+\mathcal{E}$, where $m$ is a linear operator, has received considerably less attention. When a Hilbertian framework is considered, $m \equiv m_{\beta}$ is usually assumed to be a Hilbert--Schmidt operator between $L^2$ spaces admitting an integral representation in terms of a bivariate kernel $\beta$. \cite{RamsaySilverman05} proposed to estimate $\beta$ based on minimizing the residual sum of squared norms. Motivated by signal transmission problems, \cite{Cuevasetal02} provided an estimator considering a fixed and triangular design. An estimator in terms of the Karhunen--Lo\`eve expansions of functional response and regressor was discussed in \cite{Yaoetal05}. \cite{CrambesMas13} provided asymptotic results for prediction under the FLMFR through the Karhunen--Lo\`eve expansion of the functional regressor, whereas \cite{ImaizumiKato18} derived minimax optimal rates. An estimation based on functional canonical correlation analysis was suggested in \cite{Heetal10}. The FLMFR when both response and covariate are densities was analyzed in \cite{ParkQian12}.\\

Several authors have contributed to the Goodness-of-Fit (GoF) framework for regression models, see \cite{GonzalezManteigaCrujeiras13} for a comprehensive review. The first attempts, following the ideas of \cite{BickelRosenblatt73} in scalar and multivariate  contexts, were focused on smoothing-based tests, see \cite{HardleMammen93}. Alternatively, upon the work of \cite{Durbin73}, and aimed at solving the sensitiveness of those approaches to the smoothing parameter, \cite{Stute97} proposed a GoF test based on the integrated regression function. Extending this work to the high-dimensional context, \cite{Escanciano06} proposed a GoF test, in terms of a residual marked empirical process based on projections, designed to overcome the poor empirical power inherent to the curse of dimensionality. Promoting these ideas to the FDA context, \cite{GarciaPortuguesetal14} and \cite{Cuestaetal19} derived an easily computable GoF test for the FLMSR in terms of projections. The former proposed a methodology based on the projected empirical estimator of the integrated regression function, whereas the latter considered marked empirical process indexed by a single randomly projected functional covariate, providing a more computationally efficient test.\\

In addition to the GoF proposals for the FLMSR discussed above, \cite{Delsoletal11} formulated a kernel-based test for model assumptions, whereas \cite{Buckeretal11} introduced testing procedures well-adapted for the time-variation of directional profiles. Generalized likelihood ratio tests were suggested in \cite{McLeanetal15} to test the linearity of functional generalized additive models. \cite{Staicuetal15} tested the equality of multiple group mean functions for hierarchical functional data. In the context of semi-functional partial linear model, where the scalar response is regressed on multivariate and functional covariates, \cite{AneirosPerezVieu13} tested the simple linear null hypothesis. In the FLMSR setup, a comparative study has been recently provided by \cite{Yaseminetal18}, comparing GoF tests in  \cite{HorvathReeder13}, \cite{GarciaPortuguesetal14}, \cite{McLeanetal15}, and \cite{Kongetal16}.\\

The extension of these GoF proposals to the FLMFR context is currently an open challenge. This model is being applied to a wide range of fields, such as electricity market \citep{Benatiaetal17}, biology \citep{Heetal10} or the study of lifetime patterns \citep{ImaizumiKato18}, to cite but some, hence the practical relevance of developing a GoF test for it. Testing the lack of effect, which is actually a particular case of the FLMFR, has received considerable attention: \cite{Kokoszkaetal08} proposed an FPC-based significance test within the FLMFR; \cite{Patileaetal16} introduced a kernel-based significance test consistent to nonlinear alternatives; \cite{Leeetal2020} proposed a significance test, within the FLMFR, using an extension to the functional setup of the correlation-based metric adopted in \cite{Parketal15}. Related testing approaches within the FLMFR include those of \cite{ChiouMuller07}, which addressed the development of a FPC-based residual diagnostic tool, and \cite{Gabrysetal2010}, that tested if functional residuals are independent and identically distributed (iid). Sharing the aim of the time-domain-based test in \cite{Gabrysetal2010}, \cite{Zhang2016} has recently proposed a Cram\'er--von Mises test for the functional white noise, with applications to assessing the uncorrelatedness of the residuals in FLMFR and functional autoregressive model fits, but under a frequency-domain framework, in terms of the functional periodogram previously derived in \cite{PanaretosTavakoli13}. Empirical likelihood ratio tests were formulated by \cite{Wangetal18} for concurrent models. No proposals extending the generalized likelihood ratio test approach seem to exist for the FLMFR. As a consequence, the development of GoF tests for the FLMFR, against unspecified alternatives, is an area still substantially unexplored.\\

In this paper, we propose a GoF test for the FLMFR, that is, for testing the composite null hypothesis
\begin{align*}
\mathcal{H}_0:~ m \in \mathcal{L} = \biggr\lbrace & m_{\beta}(\mathcal{X})(t) = \int_{a}^{b} \mathcal{X} (s) \beta (s,t) \,\mathrm{d}s : \beta\in L^2\left([a,b]\times[c,d]\right) \biggr\rbrace.
\end{align*}

Our methodology is based on characterizing $\mathcal{H}_0$ in terms of the integral regression operator arising from a double projection, of the functional covariate and the response, in terms of finite-dimensional functional directions. The deviation of the resulting empirical process from its expected zero mean is measured by a Cram\'er--von Mises statistic that integrates on both functional directions and is calibrated via an efficient wild bootstrap on the residuals. We show that our GoF test exhibits an adequate behavior, in terms of size and power, for the composite hypothesis, under two common scenarios: the no effects model and the FLMFR. Besides, since the test can be readily modified for the simple hypothesis $\beta\equiv0$, we compare our GoF test with the procedures from \cite{Kokoszkaetal08} and \cite{Patileaetal16}, obtaining competitive powers. As a by-product contribution, we provide a convenient hybrid approach for the estimation of $\beta$ based on LASSO \citep{Tibshirani96} regularization and linearly-constrained least-squares. The companion R package \texttt{goffda} \citep{Garcia-Portugues:goffda} implements all the methods presented in the paper and allows for replication of the real data applications.\\

The rest of this paper is organized as follows. Section \ref{sec:2} introduces the required background on FDA and the FLMFR, addressing the estimation of the regression operator and providing a brief comparative study between different estimation techniques. Section \ref{sec:3} is devoted to the theoretical, computational, and resampling aspects of the new GoF test. A comprehensive simulation study and a real data application are presented in Sections  \ref{sec:4} and \ref{sec:5}, respectively. Conclusions are drawn in Section \ref{sec:6}. Appendix \ref{sec:poofs} contains the proofs of the lemmas and the Supporting Information (SI) provides another data application.

\section{Functional data and the FLMFR}
\label{sec:2}

\subsection{Functional bases}
\label{sec:21}

Given the functional bases $\lbrace \Psi_{j}\rbrace_{j=1}^\infty$ and $\lbrace \Phi_{k}\rbrace_{k=1}^\infty$ in the separable Hilbert spaces $\mathbb{H}_1$ and $\mathbb{H}_2$, respectively, any elements $\mathcal{X} \in \mathbb{H}_1$ and $\mathcal{Y} \in \mathbb{H}_2$ can be represented as $\mathcal{X} = \sum_{j=1}^{\infty} x_j \Psi_{j} $ and $\mathcal{Y} = \sum_{k=1}^{\infty} y_k \Phi_{k}$, where $x_j = \langle \mathcal{X} , \Psi_{j} \rangle_{\mathbb{H}_1}$ and $y_k = \langle \mathcal{Y} , \Phi_{k} \rangle_{\mathbb{H}_2}$, for each $j,k \geq 1$. Typical examples are the B-splines basis (non-orthogonal piece-wise polynomial bases) or the Fourier basis. Both bases are of a deterministic nature and, despite their flexibility, usually require a larger number of elements to adequately represent a functional sample $\{\mathcal{X}_i\}_{i=1}^n$. A more parsimonious representation can be achieved by considering data-driven orthogonal bases, being the most popular choice the (empirical) Functional Principal Components (FPC) of $\{\mathcal{X}_i\}_{i=1}^n$, $\lbrace \hat{\Psi}_j\rbrace_{j=1}^n$, the eigenfunctions of the sample covariance \linebreak operator.\\

To develop the test, we will consider a $p$-truncated basis $\lbrace \Psi_{j}\rbrace_{j=1}^p$ in $\mathbb{H}_1$, corresponding to the first $p$ elements of $\lbrace \Psi_{j}\rbrace_{j=1}^\infty$. The projection of $\mathcal{X}$ on this truncated basis is denoted by $\mathcal{X}^{(p)}=\sum_{j=1}^{p} x_{j} \Psi_{j}$ and we set $\mathbf{x}_{p}:=(x_1,\ldots,x_p)$. We will also require to integrate on the functional analogue of the Euclidean $(p-1)$-sphere $\mathbb{S}^{p-1}=\{\mathbf{x}\in\mathbb{R}^{p}:\|\mathbf{x}\|=1\}$, the $(p-1)$-sphere of $\mathbb{H}_1$ on  $\lbrace \Psi_{j}\rbrace_{j=1}^\infty$ defined as $\mathbb{S}_{\mathbb{H}_1,\lbrace \Psi_{j}\rbrace_{j=1}^\infty}^{p-1} := \lbrace f=\sum_{j=1}^{p}x_{j}\Psi_{j}\in \mathbb{H}_1: \left\| f\right\|_{\mathbb{H}_1}=1 \rbrace$. The relationship between $\mathbb{S}^{p-1}$ and $\mathbb{S}_{\mathbb{H}_1,\lbrace \Psi_{j}\rbrace_{j=1}^\infty}^{p-1}$ follows easily \citep{GarciaPortuguesetal14} considering the positive semi-definite matrix  $\boldsymbol{\Psi}= \left( \langle \Psi_{j},\Psi_{\ell} \rangle_{\mathbb{H}_1} \right)_{j,\ell=1,\ldots,p}$, whose Cholesky decomposition is $\boldsymbol{\Psi}=\mathbf{P}_p'\mathbf{P}_p$. Then, the $(p-1)$-ellipsoid  $\mathbb{S}_{\boldsymbol{\Psi}}^{p-1}=\left\lbrace \mathbf{x}\in \mathbb{R}^{p}: \mathbf{x}'\boldsymbol{\Psi}\mathbf{x}=1 \right\rbrace$ is trivially isomorphic with  $\mathbb{S}_{\mathbb{H}_1,\lbrace \Psi_{j}\rbrace_{j=1}^\infty}^{p-1}$ by  $f = \sum_{j=1}^{p} x_{j}\Psi_{j}\in\mathbb{S}_{\mathbb{H}_1,\lbrace \Psi_{j}\rbrace_{j=1}^\infty}^{p-1}$ $ \mapsto \mathbf{x}_p\in\mathbb{S}_{\boldsymbol{\Psi}}^{p-1}$. Considering also the linear mapping $\mathbf{x}\in \mathbb{S}^{p-1}\mapsto  \mathbf{P}_p^{-1}\mathbf{x}\in\mathbb{S}_{\boldsymbol{\Psi}}^{p-1}$, the integration of a functional operator $\mathcal{T}$ with respect to $\gamma^{(p)} \in \mathbb{S}_{\mathbb{H}_1,\lbrace \Psi_{j}\rbrace_{j=1}^\infty}^{p-1}$ can be written as
\begin{align}
\int_{\mathbb{S}_{\mathbb{H}_1,\lbrace \Psi_{j}\rbrace_{j=1}^\infty}^{p-1}}\!\!\!\!\!\!\!\!\!\!\! \mathcal{T}\big(\gamma^{(p)} \big)\,\mathrm{d}\gamma^{(p)} = \int_{\mathbb{S}_{\boldsymbol{\Psi}}^{p-1}}\!\!\mathcal{T}\bigg( \sum_{j=1}^{p}g_{j}\Psi_{j} \bigg)\, \mathrm{d}\mathbf{g}_p =\int_{\mathbb{S}^{p-1}}\!\!\vert \mathbf{P}_p\vert^{-1} \mathcal{T}\bigg( \sum_{j=1}^{p}(\mathbf{P}_p^{-1}\mathbf{g}_p)_{j}\Psi_{j} \bigg)\, \mathrm{d}\mathbf{g}_p,
\label{eq_1}
\end{align}
where $\left(\mathbf{P}_p^{-1} \mathbf{g}_p \right)_{j}$ denotes the $j$-th component of the vector $\mathbf{P}_p^{-1} \mathbf{g}_p$ and $\mathbf{g}_p$ is the vector of coefficients of $\gamma^{(p)}$ in the $p$-truncated basis. If the basis is orthonormal, then $\boldsymbol{\Psi}$ and $\mathbf{P}_p$ are the identity matrices of order $p$, denoted as $\mathbf{I}_p$, and $\mathbf{g}_p \in \mathbb{S}^{p-1}$ without any transformation. Clearly, an analogous development can be established for $\mathbb{S}_{\mathbb{H}_2,\lbrace \Phi_{k}\rbrace_{k=1}^\infty}^{q-1}$ by means of   $\boldsymbol{\Phi} = \left( \langle \Phi_k, \Phi_\ell \rangle_{\mathbb{H}_2} \right)_{k,\ell=1,\ldots,q}$ where $\lbrace \Phi_{k}\rbrace_{k=1}^q$ is a $q$-truncated basis in $\mathbb{H}_2$.

\subsection{The FLMFR}
\label{sec:22}

We consider the context of functional regression with $\mathbb{H}_2$-valued functional response $\mathcal{Y}$ and $\mathbb{H}_1$-valued functional covariate $\mathcal{X}$:
\begin{align}
\mathcal{Y}=m(\mathcal{X})+\mathcal{E},\label{eq:modreg}
\end{align}
where the regression operator is defined as $m({\scriptstyle{\mathcal{X}}})=\mathbb{E} \left[ \mathcal{Y}|\mathcal{X}={\scriptstyle{\mathcal{X}}} \right]$ and the $\mathbb{H}_2$-valued error is such that $\mathbb{E} \left[ \mathcal{E} | \mathcal{X} \right] = 0$. Within this setting, we assume that $\mathcal{X}$ and $\mathcal{Y}$ are already centered so there is no need for an intercept term in \eqref{eq:modreg}. Particularly, we consider $L^2$ spaces and assume, in what follows, that $\mathcal{X} \in \mathbb{H}_1 = L^2 \left([a,b] \right)$ and $\mathcal{Y} \in \mathbb{H}_2 = L^2 \left([c,d]\right)$, unless otherwise explicitly mentioned.\\

In this context, the simplest parametric model is the FLMFR, in which the regression operator $m:\mathbb{H}_1 \longrightarrow \mathbb{H}_2$ is usually assumed to be a Hilbert--Schmidt integral operator, i.e., $m$ admits an integral representation $m_\beta$ given by a bivariate kernel $\beta\in \mathbb{H}_1\otimes\mathbb{H}_2=L^2([a,b]\times[c,d])$ as follows:
\begin{align}
m_{\beta} (\mathcal{X})(t)=\int_{a}^{b} \beta(s,t)\mathcal{X}(s)\,\mathrm{d}s, \quad t \in [c,d].
\label{linear_op}
\end{align}

In particular, the Hilbert--Schmidt condition directly implies that $m$ is a compact operator, that is, $\beta$ can be decomposed in terms of the tensor product of any pair of bases in $\mathbb{H}_1$ and $\mathbb{H}_2$, since such tensor product constitutes a basis on the space of Hilbert--Schmidt operators. As a consequence,
\begin{align}
\beta = \sum_{j=1}^\infty \sum_{k=1}^\infty b_{jk} (\Psi_j \otimes \Phi_k),\ b_{jk} = \frac{\left\langle \beta, \Psi_j\otimes\Phi_k\right\rangle_{\mathbb{H}_1\otimes\mathbb{H}_2}}{\|\Psi_j\|_{\mathbb{H}_1}^2\|\Phi_k\|_{\mathbb{H}_2}^2},\!\!\label{eq:beta}
\end{align}
with $j,k \geq 1$. For convenience, we denote the linear integral operator in \eqref{linear_op} by $\langle\langle\cdot,\star\rangle\rangle$, defined as
$\langle\langle\cdot,\star\rangle\rangle\colon \mathbb{H}_1\times (\mathbb{H}_1 \otimes \mathbb{H}_2) \longrightarrow \mathbb{H}_2$, $ \langle\langle \mathcal{X},\beta \rangle\rangle(t):=\langle \mathcal{X},\beta(\cdot, t)\rangle_{\mathbb{H}_1}$.
Therefore, the FLMFR from \eqref{eq:modreg}--\eqref{linear_op} can be succinctly denoted as
\begin{align}
\mathcal{Y}= \langle\langle \mathcal{X},\beta \rangle\rangle+\mathcal{E}.\label{eq:moddoub}
\end{align}
Bearing in mind that $\mathcal{X} = \sum_{j=1}^{\infty} x_j \Psi_j$ and $\mathcal{Y} = \sum_{k=1}^{\infty} y_k \Phi_k$, then
\begin{align}
\langle\langle \mathcal{X},\beta \rangle\rangle = \bigg\langle \bigg\langle \sum_{j=1}^{\infty} x_{j}\Psi_{j}, \sum_{\ell=1}^{\infty}\sum_{k=1}^{\infty} b_{\ell k}(\Psi_{\ell}\otimes \Phi_{k})\bigg\rangle\bigg\rangle = \sum_{j=1}^{\infty}\sum_{\ell=1}^{\infty} \sum_{k=1}^{\infty} b_{\ell k}x_{j} \langle \Psi_{j},\Psi_{\ell} \rangle_{\mathbb{H}_1} \Phi_{k},
\label{B}
\end{align}
with $\langle \Psi_{j},\Psi_{\ell} \rangle_{\mathbb{H}_1}=\delta_{j\ell}$, $j,\ell \geq 1$, for orthonormal bases. From \eqref{B} and $\mathcal{E} = \sum_{k=1}^{\infty} e_k \Phi_k$,
\begin{align*}
y_k = \sum_{j=1}^{\infty} \sum_{\ell=1}^{\infty} b_{\ell k}x_{j} \langle \Psi_{j},\Psi_{\ell} \rangle_{\mathbb{H}_1} + e_k, \ k \geq 1.
\end{align*}

This (infinite) linear model is usually approached by projecting the variables in the truncated bases $\lbrace \Psi_{j}\rbrace_{j=1}^p$ and $\lbrace \Phi_{k}\rbrace_{k=1}^q$ \citep[Chapter 16]{RamsaySilverman05}, obtaining the $(p,q)$-truncated population version
\begin{align}
y_k =  \sum_{j=1}^{p} \sum_{\ell=1}^{p} b_{\ell k}x_{j} \langle \Psi_{j},\Psi_{\ell} \rangle_{\mathbb{H}_1} + e_k,\  k=1,\ldots,q.\label{eq:pop}
\end{align}
Note that an equivalent way of expressing \eqref{eq:pop} is $\mathcal{Y}^{(q)}=\left\langle\left\langle \mathcal{X}^{(p)},\beta^{(p,q)} \right\rangle\right\rangle+\mathcal{E}^{(q)}$,
where $\beta^{(p,q)}$ is the projection of \eqref{eq:beta} into $\{\Psi_j \otimes \Phi_k\}_{j,k=1}^{p,q}$.\\

Now, given an iid centered sample $\{(\mathcal{X}_{i},\mathcal{Y}_{i})\}_{i=1}^n$ such that $\mathcal{Y}_{i}=\langle\langle\mathcal{X}_{i},\beta \rangle\rangle+\mathcal{E}_{i}$, the sample version of \eqref{eq:pop} is expressed in matrix form as
\begin{align}
\mathbf{Y}_{q} = \mathbf{X}_{p} \boldsymbol{\Psi} \mathbf{B}_{p,q}+\mathbf{E}_q,\label{eq:samp}
\end{align}
where $\mathbf{Y}_{q}$ and $\mathbf{E}_{q}$ are the $n\times q$ matrices with the coefficients of $\{\mathcal{Y}_i\}_{i=1}^n$ and $\{\mathcal{E}_i\}_{i=1}^n$, respectively, on $\lbrace \Phi_{k}\rbrace_{k=1}^q$, $\mathbf{X}_{p}$ is the $n\times p$ matrix of coefficients of $\{\mathcal{X}_i\}_{i=1}^n$ on $\lbrace \Psi_{j}\rbrace_{j=1}^p$, and $\mathbf{B}_{p,q}$ is the $p\times q$ matrix of unknown coefficients on $\lbrace \Psi_{j}\otimes\Phi_k\rbrace_{j,k=1}^{p,q}$. Observe that these matrices are centered by columns and hence the model does not have an intercept. Clearly, due to the form of \eqref{eq:samp}, estimators for $\beta$ in \eqref{eq:beta} readily follow from the linear model theory. We discuss them next, focusing exclusively on orthonormal bases. This can be done without loss of generality; just replace $\mathbf{X}_{p}$ by $\breve{\mathbf{X}}_{p}:=\mathbf{X}_{p} \boldsymbol{\Psi}$ subsequently for non-orthonormal bases.

\subsection{Model estimation}
\label{sec:23}

FPCR considers in \eqref{eq:samp} the data-driven bases given by the (empirical) FPC $\lbrace \hat\Psi_{j}\rbrace_{j=1}^p$ and $\lbrace \hat\Phi_{k}\rbrace_{k=1}^q$ of $\{\mathcal{X}_i\}_{i=1}^n$ and $\{\mathcal{Y}_i\}_{i=1}^n$, respectively, where $p,q\leq n$. The estimator of $\beta$ is then defined as the least-squares estimator of the $(p,q)$-truncated model given in \eqref{eq:pop} and \eqref{eq:samp}:
\begin{align*}
\hat{\mathbf{B}}_{p,q} =\mathrm{arg}\min_{\mathbf{B}_{p,q}}\left\| \mathbf{Y}_q-\mathbf{X}_{p}  \mathbf{B}_{p,q} \right\|^{2}=\mathrm{arg}\min_{\beta^{(p,q)}}\sum_{i=1}^n\left\| \mathcal{Y}_i^{(q)}-\big\langle\big\langle\mathcal{X}_i^{(p)},\beta^{(p,q)}\big\rangle\big\rangle \right\|^{2}.
\end{align*} 
Clearly, least-squares estimation gives $\hat{\mathbf{B}}_{p,q} = \left(\mathbf{X}_{p}' \mathbf{X}_{p}  \right)^{-1} \mathbf{X}_{p}' \mathbf{Y}_{q}$, with $(\hat{\mathbf{B}}_{p,q})_{jk} = \hat{b}_{jk}$, $j=1,\ldots,p$, $k=1,\ldots,q$. The estimator of $\beta^{(p,q)}$ is then $\hat{\beta}^{(p,q)} =  \sum_{j=1}^{p}\sum_{k=1}^{q} \hat{b}_{jk}(\hat\Psi_{j}\otimes \hat\Phi_{k})$.\\

The estimator $\hat{\beta}^{(p,q)}$ critically depends on $(p,q)$, hence an automatic data-driven selection of $(p,q)$ is of most practical interest. A possibility is to extend the predictive cross-validation criterion from \cite{PredaSaporta05} to the FLMFR context, at expenses of a likely high computational cost (cross-validation on two indexes). Alternatives based on the generalized cross-validation procedure \citep{Cardotetal03} or a stepwise model selection approach based on the BIC criterion could be studied, but neither the degrees of freedom or the likelihood function are immediate to estimate in the FLMFR setup. A feasible possibility, though not regression-driven, is to select $p$ and $q$ as the minimum number of components associated with a certain proportion of Explained Variance ($\mathrm{EV}_p$ and $\mathrm{EV}_q$), e.g., such that $\mathrm{EV}_p=\mathrm{EV}_q=0.99$. This simple rule provides an initial selection which can be subsequently improved.\\

Regularization techniques provide an estimation alternative that, due to their flexibility and efficient computational implementations \citep{Friedman2010}, have been remarkably popular in the last decades. The so-called elastic-net regularization of $\mathbf{B}_{p,q}$ gives the estimator
\begin{align*}
\hat{\mathbf{B}}_{p,q}^{(\lambda)}  =  \mathrm{arg}\min_{\mathbf{B}_{p,q}}\biggr\{\frac{1}{2n}  \sum_{i=1}^{n} \left\|\left(\mathbf{Y}_q \right)_i - \left(\mathbf{X}_{p}  \mathbf{B}_{p,q}\right)_i
\right\|^{2} + \lambda \biggr[ \frac{1-\alpha}{2} \| \mathbf{B}_{p,q} \|_{\mathrm{F}}^{2} + \alpha  \sum_{j=1}^{p} \left\|\left(\mathbf{B}_{p,q}\right)_j\right\|_2 \biggr]
\biggr\}, \end{align*} 
where $\lambda\geq0$ is the penalty parameter, $\alpha\in [0,1]$, $\|\cdot\|_{\mathrm{F}}$ is the Frobenius norm, and $(\mathbf{A})_i$ stands for the $i$-th row of the matrix $\mathbf{A}$. If $\lambda=0$, then we the usual FPCR follows. Cases $\alpha=0$ and $\alpha=1$ correspond to ridge (henceforth denoted as FPCR-L2) and LASSO (FPCR-L1) regression, respectively. The former does a global penalization in all the entries of $\mathbf{B}_{p,q}$, whereas the latter applies a row-wise penalization that effectively zeroes full rows, hence removing predictors. Therefore, the key advantage of the FPCR-L1 is that it enables variable selection: $p$ and $q$ are initially fixed but only $\tilde{p} \leq p$ components are selected. On the other hand, FPCR-L2 exhibits an important advantage when employed within the bootstrap algorithm to be described in Section \ref{sec:34}: the estimation $\hat{\mathbf{Y}}_{q} = \mathbf{X}_{p}  \hat{\mathbf{B}}_{p,q}^{(\lambda)}$ can be re-expressed as $\hat{\mathbf{Y}}_{q} = \mathbf{H}^{(\lambda)} \mathbf{Y}_q,$  where $\mathbf{H}^{(\lambda)} = \mathbf{X}_{p}  \left(\mathbf{X}_{p}' \mathbf{X}_{p}  + \lambda \mathbf{I}_p \right)^{-1} \mathbf{X}_{p}'$ is the hat matrix for the FPCR-L2 estimator. The lack of an analogous result for the FPCR-L1 estimator notably increases the bootstrapping cost. Finally, note that $\lambda$ can be selected with reasonable efficiency through leave-one-out cross-validation ($\hat\lambda_{\mathrm{CV}}$), as implemented in \cite{Friedman2010}.\\

As a way to exploit the advantages of both FPCR-L1 and FPCR-L2, we propose a hybrid approach, termed FPCR-L1-selected (FPCR-L1S) estimator, which firstly implements FPCR-L1 for variable selection, and then performs FPCR estimation with the predictors selected by FPCR-L1 (see Remark \ref{rem:7} on variable selection by FPCR-L1). Therefore, FPCR-L1S has a hat matrix that is very convenient for the latter bootstrap algorithm:
\begin{align}
\mathbf{H}_\mathrm{C}^{(\lambda)} = \Tilde{\mathbf{X}}_{\tilde{p}}  \left(\Tilde{\mathbf{X}}_{\tilde{p}}' \Tilde{\mathbf{X}}_{\tilde{p}}  \right)^{-1} \Tilde{\mathbf{X}}_{\tilde{p}}',
\label{hat_matrix_FPCR-L2-C}
\end{align}
where $\Tilde{\mathbf{X}}_{\tilde{p}}$ is the matrix of the coefficients of the $\tilde{p}$ selected predictors (which can be \emph{non-consecutive} FPC). This variable selection is a crucial advantage, as clearly the number of FPC for representing $\mathcal{X}$ and $\mathcal{Y}$ up to a certain EV might not correspond with the best selection of $(p,q)$ for the estimation of $\mathbf{B}_{p,q}$. We denote the scores of the FPCR-L1S estimator as $\hat{\mathbf{B}}^{(\lambda),\mathrm{C}}_{\tilde{p},q}$.

\subsection{Comparative study of estimators}
\label{sec:24}

A succinct simulation study is conducted for comparing the performance of the four estimators previously described. We used the following common settings: the functional covariates $\left\lbrace \mathcal{X}_i \right\rbrace_{i=1}^{n}$ are  centered and valued in $[0,1]$, the functional errors $\left\lbrace \mathcal{E}_i \right\rbrace_{i=1}^{n}$ are valued  in $[2,3]$ (both intervals were discretized in $101$ equispaced grid points), the sample size is $n = 100$, and $1,000$ Monte Carlo replicates were considered. The simulation scenarios are collected in Table \ref{tab:scenarios} and have the following descriptions:

\begin{itemize}
	\item \textbf{CM.} Based on the process used in \cite{CrambesMas13}, where $\mathcal{X} (s) = \sum_{j=1}^{50} \lambda_j \varepsilon_{j} \Psi_j (s)$, $\varepsilon_j \sim \mathcal{N}(0,2^2)$, with $\lambda_j = (\pi^2 (j-\frac{1}{2})^2)^{-1}$ and $\Psi_j (s) = \sqrt{2} \sin ((j-0.5) \pi s)$, for each $j \geq 1$ and $s \in [0,1]$.
	
	\item \textbf{BM.} Brownian motion with standard deviation equal to $0.15$.
	
	\item \textbf{IK.} Based on the process used in \cite{ImaizumiKato18}. Functional covariates are given by $\mathcal{X}(s)  = \sum_{j=1}^{50} j^{-7/4} U_{j} \Psi_j(s),~ U_j \sim \mathcal{U}(-\sqrt{5},\sqrt{5})$, with $\Psi_1(s) \equiv 1$ and $\Psi_j (s) = \sqrt{2} \cos (j\pi s) $, for each $j \geq 1$ and $s \in [0,1]$. Functional errors are given by $\mathcal{E}(t)  = \sum_{j=1}^{50} j^{-4/5} \varepsilon_j \Psi_j(t),~ \varepsilon_j \sim \mathcal{N}(0,1.5^2)$, for each $j \geq 1$ and $t \in [2,3]$.
	
	\item \textbf{GP.} Gaussian process with covariance function $\Sigma (s_1,s_2)= 6^2 \exp(-|s_1-s_2|/0.2)$.
	
	\item \textbf{OU.} Ornstein--Uhlenbeck process  with unitary drift and stationary standard deviation equal to $0.35$.
\end{itemize}

\begin{table}[htpb!]
	\iffigstabs
	\footnotesize
	\centering
	\begin{tabular}{cccc}
		\toprule
		Scenario  & Kernel $\beta(s,t)$ & $\mathcal{X}(s)$ &  $\mathcal{E}(t)$
		\\
		\midrule
		S1 & $(s-a)^2 + (t-c)^2$ & CM & BM \\ \midrule
		S2 & $2\left[\sin(6 \pi (s-a)) + \cos(6\pi (t-c))\right]$ & GP & OU  \\\midrule
		\multirow{2}{*}{S3} & $\beta(s,t) =  \sum_{j=1}^{\infty} \sum_{k=1}^{\infty} b_{j,k} \Phi_j (s) \Psi_k(t)$ with $b_{jk} = 0$ if $j,k\leq 4$, & \multirow{2}{*}{ IK} & \multirow{2}{*}{ IK} \\
		&  $b_{jk} = 6 (-1)^{j+k} \left(j - 4 \right)^{-12/5} \left(k - 4 \right)^{-1/4}$ otherwise  &  &  \\ \bottomrule
	\end{tabular}
	\fi
	\caption{\small{Summary of the simulated scenarios. 
			\label{tab:scenarios}}}
\end{table}

Table \ref{table:01} shows the averaged errors $\big\|\beta-\hat\beta^{(p,q)}\big\|_{\mathbb{H}_1\otimes\mathbb{H}_2}$ of all estimators for $p= 2, 5, 10, 25, 50$ and $q= 1, 5, 10$, with $\lambda$ set as $\hat\lambda_{\mathrm{CV}}$. We summarize next the conclusions: 
\begin{itemize}
	\item There is a weak dependency on $q$: parameters $(p,q)$ do not play a symmetric role \citep{RamsaySilverman05}. Nonetheless, the influence of $q$ is more prevalent in S2 and S3, inasmuch as an amount of EV has still to be captured.
	
	\item When $p$ is excessively large, errors skyrocket for FPCR and FPCR-L2, in contrast with FPCR-L1 and FPCR-L1S. This is clearly observed in S1 (low variability and a linear kernel), since the model begins to become promptly overfitted ($\text{EV}_{p = 2} > 0.99$ and $\text{EV}_{q = 1} = 0.98$) and the effective variable selection of FPCR-L1 and FPCR-L1S is clearly manifested ($\bar{\tilde{p}} / p < 0.05$ as $p$ increases).
	
	\item S2 (high variability and an egg-carton-shape-like kernel) illustrates the situation in which the functional samples are not properly represented with few FPC ($\text{EV}_{p = 10} < 0.95$). Even though errors are smaller than in S1 (overfitting is mitigated, $\bar{\tilde{p}} / p \simeq 0.25$ as $p$ increases), FPCR-L1 (mainly) and FPCR-L1S provide more precise estimations. FPCR slightly outperforms the rest of estimators for small values of $(p,q)$.
	
	\item A sensible choice of $(p,q)$ for representing the functional samples might not be so for estimating $\beta$. This is illustrated in S3: even though $\mathcal{X}$ and $\mathcal{Y}$ are smoother than in S2, $\bar{\tilde{p}}$ is not much smaller, since the first components are not informative. The number of selected FPC for FPCR-L1 and FPCR-L1S is drastically reduced  for large values of $(p,q)$ ($\bar{\tilde{p}} / p < 0.05$, when $p = 50$ and $q = 10$), since non-consecutive FPC are allowed to be selected, removing the noise from estimating the first null components.
\end{itemize}

All in all, FPCR-L1 outperforms FPCR-L1S, yet both performances are markedly better than the FPCR and FPCR-L2 ones. Because of this and the key computational advantage the explicit hat matrix \eqref{hat_matrix_FPCR-L2-C} delivers, we will adopt FPCR-L1S as our reference estimator. 

\begin{table}[h!]
	\iffigstabs
	\setlength{\tabcolsep}{1.5pt}
	\renewcommand{\arraystretch}{0.9}
	\centering
	\footnotesize
	\begin{tabular}{cc|ccc|ccc|ccc}
		\toprule
		\multicolumn{2}{c|}{Scenario} & \multicolumn{3}{c|}{S1} & \multicolumn{3}{c|}{S2} & \multicolumn{3}{c}{S3} \\
		\midrule
		\multicolumn{2}{c|}{$q$ ($\text{EV}_q$)} & 1 ({\scriptsize $98\%$}) & 5 ({\scriptsize $>99\%$}) & 10 ({\scriptsize $>99\%$})    & 1 ({\scriptsize $92\%$}) & 5 ({\scriptsize $>99\%$}) & 10 ({\scriptsize $>99\%$}) &  1 ({\scriptsize $38\%$}) & 5 ({\scriptsize $86\%$}) & 10 ({\scriptsize $93\%$}) \\
		\midrule 
		& $\text{EV}_p$ & \multicolumn{3}{c|}{$>99\%$} & \multicolumn{3}{c|}{$68.69\%$} &
		\multicolumn{3}{c}{$96.66\%$}  \\
		\cmidrule{3-11}
		& $\bar{\tilde{p}}$ & {\scriptsize $1.0\,(0.13)$} & {\scriptsize $1.0\,(0.13)$} & {\scriptsize $1.0\,(0.13)$}  &  {\scriptsize $1.8\,(0.40)$ } & {\scriptsize $2.0\,(0.14)$} & {\scriptsize $2.0\,(0.14)$ }  &  {\scriptsize $1.0\,(0.00)$} & {\scriptsize $1.0\,(0.00)$} & {\scriptsize $1.0\,(0.00)$ }   \\
		\midrule
		\multirow{4}{*}{$p=2$ \hspace{-0.1cm}} & {\scriptsize FPCR } & $0.303$ & $0.296$ & $0.296$ & $\mathbf{1.438}$ & $\mathbf{1.418}$ & $\mathbf{1.418}$ & $\mathbf{21.382}$ & $21.386$ & $21.387$ \\
		& {\scriptsize L1} & $0.216$ & $0.216$ & $0.216$ & $1.438$ & $1.425$ & $1.425$ & $21.385$ & $\mathbf{21.385}$ & $\mathbf{21.385}$ \\
		& {\scriptsize L2 } & $0.300$ & $0.291$ & $0.291$ & $1.438$ & $1.423$ & $1.423$ & $21.385$ & $\mathbf{21.385}$ & $\mathbf{21.385}$ \\
		& {\scriptsize L1S} & $\mathbf{0.204}$ & $\mathbf{0.203}$ & $\mathbf{0.203}$ & $1.438$ & $1.418$ & $1.418$ & $21.385$ & $\mathbf{21.385}$ & $\mathbf{21.385}$ \\
		\midrule 
		& $\text{EV}_p$ &  \multicolumn{3}{c|}{$>99\%$}  & \multicolumn{3}{c|}{$87.77\%$} &\multicolumn{3}{c}{$98.33\%$}  \\
		\cmidrule{3-11}
		& $\bar{\tilde{p}}$ & {\scriptsize $1.1\,(0.27)$} & {\scriptsize $1.1\,(0.28)$} & {\scriptsize $1.1\,(0.28)$ }  &  {\scriptsize $4.8\,(0.47)$ } & {\scriptsize  $4.4\,(0.70)$ } & {\scriptsize  $4.4\,(0.70)$ }  & {\scriptsize $3.1\,(1.40)$ } & {\scriptsize $1.5\,(0.83)$} & {\scriptsize $1.5\,(0.81)$} \\
		\midrule
		\multirow{3}{*}{$p=5$ \hspace{-0.1cm}}& {\scriptsize FPCR} & 2.461 & 2.660 & 2.670 & $\mathbf{1.418}$ & $\mathbf{1.303}$ & $\mathbf{1.304}$ & $\mathbf{10.204}$ & $\mathbf{6.696}$ & $\mathbf{6.738}$ \\
		& {\scriptsize L1 } & $\mathbf{0.239}$ & $\mathbf{0.242}$ & $\mathbf{0.243}$ & $1.418$ & $1.326$ & $1.326$ & $10.299$ & $9.182$ & $9.256$ \\
		& {\scriptsize L2 } & $2.161$ & $2.316$ & $2.324$ & $1.418$ & $1.308$ & $1.308$ & $10.417$ & $10.221$ & $10.335$ \\
		& {\scriptsize L1S} & $0.308$ & $0.323$ & $0.323$ & $1.418$ & $1.307$ & $1.307$ & $10.230$ & $\mathbf{6.711}$ & $\mathbf{6.716}$ \\
		\midrule 
		& $\text{EV}_p$ & \multicolumn{3}{c|}{$>99\%$} & \multicolumn{3}{c|}{$94.46\%$} &
		\multicolumn{3}{c}{$>99\%$}  \\
		\cmidrule{3-11}
		& $\bar{\tilde{p}}$ & {\scriptsize $1.1\,(0.45)$} & {\scriptsize $1.1\,(0.42)$} & {\scriptsize $1.1\,(0.42)$}  &  {\scriptsize $8.1\,(1.28)$ } & {\scriptsize $8.9\,(1.00)$ } & {\scriptsize $8.8\,(1.00)$ }  &  {\scriptsize $5.1\,(2.52)$ } & {\scriptsize $1.9\,(1.32)$} & {\scriptsize $1.9\,(1.28)$ }   \\
		\midrule
		\multirow{4}{*}{$p=10$ \hspace{-0.1cm}} & {\scriptsize FPCR } & $15.297$ & $16.411$ & $16.461$ & $\mathbf{1.416}$ & $0.504$ & $0.507$ & $9.643$ & $14.313$ & $15.342$ \\
		& {\scriptsize L1} & $\mathbf{0.404}$ & $\mathbf{0.407}$ & $\mathbf{0.408}$ & $1.416$ & $0.547$ & $0.548$ & $\mathbf{8.981}$ & $8.782$ & $8.868$ \\
		& {\scriptsize L2} & $13.354$ & $14.194$ & $14.236$ & $1.416$ & $\mathbf{0.503}$ & $\mathbf{0.506}$ & $9.348$ & $12.468$ & $12.912$ \\
		& {\scriptsize L1S} & $1.193$ & $1.185$  & $1.186$ & $1.416$ & $0.507$ & $0.509$ & $9.175$ & $\mathbf{6.960}$ & $\mathbf{6.978}$ \\
		\midrule 
		& $\text{EV}_p$ & \multicolumn{3}{c|}{$>99\%$} & \multicolumn{3}{c|}{$98.37\%$} &
		\multicolumn{3}{c}{$>99\%$}  \\
		\cmidrule{3-11}
		& $\bar{\tilde{p}}$ & {\scriptsize $1.2\,(0.69)$} & {\scriptsize $1.2\,(0.67)$}  &  {\scriptsize  $1.2\,(0.67)$ } & {\scriptsize $11.5\,(3.22)$}  & {\scriptsize  $11.7\,(3.02)$} & {\scriptsize $11.7\,(3.01)$ }  &  {\scriptsize $5.9\,(3.75)$ } & {\scriptsize $2.0\,(1.80)$ } & {\scriptsize $1.70\,(0.45)$}   \\
		\midrule
		\multirow{4}{*}{$p=25$ \hspace{-0.1cm}} & {\scriptsize FPCR } & $164.917$ & $176.286$ & $176.757$ & $1.419$ & $1.271$ & $1.291$ & $36.794$ & $111.324$ & $119.420$ \\
		& {\scriptsize  L1}  & $\mathbf{2.006}$ & $\mathbf{2.004}$ & $\mathbf{1.986}$ & $\mathbf{1.416}$ & $\mathbf{0.622}$ & $\mathbf{0.622}$ & $\mathbf{10.358}$ & $\mathbf{10.383}$ & $\mathbf{10.290}$ \\
		& {\scriptsize L2}  & $142.442$ & $150.485$ & $150.857$ & $1.419$ & $1.222$ & $1.241$ & $26.367$ & $52.843$ & $53.747$ \\
		& {\scriptsize  L1S} & $9.549$ & $10.505$ & $10.435$ & $1.417$ & $0.936$ & $0.943$ & $16.720$ & $15.679$ & $15.310$ \\
		\midrule
		& $\text{EV}_p$ & \multicolumn{3}{c|}{$>99\%$} & \multicolumn{3}{c|}{$99\%$} &
		\multicolumn{3}{c}{$>99\%$}  \\
		\cmidrule{3-11}
		& $\bar{\tilde{p}}$ & {\scriptsize $1.5\,(1.74)$} & {\scriptsize $1.4\,(1.52)$} & {\scriptsize $1.4\,(1.48)$}  &  {\scriptsize  $13.2\,(5.50)$ } & {\scriptsize  $13.6\,(5.10)$ } & {\scriptsize  $13.6\,(5.11)$ }  &  {\scriptsize $6.8\,(5.30)$ } & {\scriptsize $2.2\,(2.57)$ } & {\scriptsize $2.2\,(2.60)$}   \\
		\midrule
		\multirow{4}{*}{$p=50$ \hspace{-0.1cm}} & {\scriptsize FPCR } & $1231.590$ & $1313.864$ & $1317.221$ & $1.445$ & $3.596$ & $3.654$ & $220.034$ & $680.661$ & $729.409$ \\
		& {\scriptsize L1} & $\mathbf{19.933}$ & $\mathbf{17.903}$ & $\mathbf{17.703}$ & $\mathbf{1.418}$ & $\mathbf{0.856}$ & $\mathbf{0.852}$ & $\mathbf{20.103}$ & $\mathbf{20.621}$ & $\mathbf{19.626}$ \\
		& {\scriptsize L2} & $1045.237$ & $1098.301$ & $1100.604$ & $1.444$ & $3.456$ & $3.510$ & $135.310$ & $212.072$ & $203.919$ \\
		& {\scriptsize L1S} & $92.410$ & $92.469$ & $91.647$ & $1.429$ & $2.097$ & $2.105$ & $60.360$ & 73.110 & $68.900$ \\ 		
		\bottomrule
	\end{tabular}
	\caption{\small{Averaged $L^2$ estimation errors. The average number (sd in parentheses) of selected FPC with FPCR-L1 and FPCR-L1S is denoted as $\bar{\tilde{p}}$. Boldfaces denote the errors that are not significantly larger than the smallest (on each block), according to a $95\%$-confidence paired $t$-test.}\label{table:01}}
	\fi
\end{table}

\section{A GoF test for the FLMFR}
\label{sec:3}

\subsection{Derivation of the test statistic}
\label{sec:31}

Our aim is to verify whether the relation between the functional response and predictor can be explained by the FLMFR in \eqref{B}, that is, to test the composite null hypothesis
\begin{align*}
\mathcal{H}_0:m \in \mathcal{L} = \left\lbrace  \langle \langle \cdot, \beta \rangle \rangle:\beta\in \mathbb{H}_1\otimes\mathbb{H}_2\right\rbrace
\end{align*}
against an unspecified alternative hypothesis $\mathcal{H}_1: \mathbb{P} \left( m \not\in \mathcal{L} \right) > 0$. Note that $\mathcal{H}_0$ is equivalent to $\mathcal{H}_0: m(\cdot) = \langle \langle \cdot, \beta \rangle \rangle$, where the equality holds for some \emph{unknown} $\beta\in\mathbb{H}_1\otimes \mathbb{H}_2$.

The following lemmas give the characterization of $\mathcal{H}_0$ in terms of the one-dimensional projections of the response and the predictor.

\begin{lemma}[$\mathcal{H}_0$ characterization]
	\label{lemma:1}
	Let $\mathcal{X}$ and $\mathcal{Y}$ be $\mathbb{H}_1$- and $\mathbb{H}_2$-valued random variables, respectively, and $\beta \in \mathbb{H}_{1}\otimes \mathbb{H}_{2}$. Then, the following statements are equivalent:
	\begin{enumerate}[label=\roman*., ref=\textit{\roman*}]
		
		\item $\mathcal{H}_0$ holds, that is, $m \left({\scriptstyle\mathcal{X}} \right) = \langle \langle {\scriptstyle\mathcal{X}} , \beta \rangle \rangle$, $\forall {\scriptstyle\mathcal{X}} \in \mathbb{H}_1$.\label{lemma:1:1}
		
		\item $\mathbb{E} \left[ \mathcal{Y} - \langle \langle \mathcal{X} , \beta \rangle \rangle | \mathcal{X} = {\scriptstyle{\mathcal{X}}}\right] = 0$, for almost every (a.e.) $ {\scriptstyle{\mathcal{X}}} \in \mathbb{H}_1$.\label{lemma:1:2}
		
		\item $\mathbb{E} \left[ \mathcal{Y} - \langle \langle \mathcal{X} , \beta \rangle \rangle | \langle \mathcal{X}, \gamma_{\mathcal{X}} \rangle_{\mathbb{H}_1} = u\right] = 0$, for a.e. $u \in \mathbb{R}$, $\forall \gamma_{\mathcal{X}} \in \mathbb{S}_{\mathbb{H}_1}$.\label{lemma:1:3}
		
		\item $\mathbb{E} \left[ \langle \mathcal{Y} - \langle \langle \mathcal{X} , \beta \rangle \rangle, \gamma_{\mathcal{Y}} \rangle_{\mathbb{H}_2} | \langle \mathcal{X}, \gamma_{\mathcal{X}} \rangle_{\mathbb{H}_1} = u\right] = 0$ almost surely (a.s.), for a.e. $u \in \mathbb{R}$ and $\forall \gamma_{\mathcal{X}} \in \mathbb{S}_{\mathbb{H}_1}, \gamma_{\mathcal{Y}} \in \mathbb{S}_{\mathbb{H}_2}$.\label{lemma:1:4}
		
		\item $\mathbb{E} \left[ \langle \mathcal{Y} - \langle \langle \mathcal{X}, \beta \rangle \rangle, \gamma_{\mathcal{Y}} \rangle_{\mathbb{H}_2} \mathbbm{1}_{\left\{\langle \mathcal{X}, \gamma_{\mathcal{X}} \rangle_{\mathbb{H}_1} \leq u\right\}}\right] = 0$ a.s., for a.e. $u \in \mathbb{R}$ and $\forall \gamma_{\mathcal{X}} \in \mathbb{S}_{\mathbb{H}_1}, \gamma_{\mathcal{Y}} \!\in\! \mathbb{S}_{\mathbb{H}_2}$.\hspace*{-1cm}\label{lemma:1:7}
		
	\end{enumerate}
\end{lemma}

\begin{lemma}[$\mathcal{H}_0$ characterization on finite-dimensional directions]\label{lemma:2}
	Within the setting of Lemma \ref{lemma:1}, let $\{\Psi_j\}_{j=1}^\infty$ and $\{\Phi_k\}_{k=1}^\infty$ be bases of $\mathbb{H}_1$ and $\mathbb{H}_2$, respectively. Then, the previous statement \ref{lemma:1:7} is equivalent to
	\begin{enumerate}[label=\roman*'., ref=\textit{\roman*'}]
		\setcounter{enumi}{4}
		\item $\mathbb{E} \left[ \langle \mathcal{Y} - \langle \langle \mathcal{X}, \beta \rangle \rangle, \gamma_{\mathcal{Y}} \rangle_{\mathbb{H}_2} \mathbbm{1}_{\left\{\langle \mathcal{X}, \gamma_{\mathcal{X}} \rangle_{\mathbb{H}_1} \leq u\right\}}\right] = 0$, for a.e. $u \in \mathbb{R}$, $\forall \gamma_{\mathcal{X}} \in \mathbb{S}_{\mathbb{H}_1,\{\Psi_j\}_{j=1}^\infty}^{p-1}, \gamma_{\mathcal{Y}} \in \mathbb{S}_{\mathbb{H}_2,\{\Phi_k\}_{k=1}^\infty}^{q-1}$, and for all $p,q \geq 1$. \label{lemma:1:8}
	\end{enumerate}
	Hence, $\mathcal{H}_0$ holds if and only if \ref{lemma:1:8} is satisfied. In addition, the former statements \ref{lemma:1:3}--\ref{lemma:1:4} are equivalent to their iii'--iv' analogues.
\end{lemma}

We use the characterization given by \ref{lemma:1:7} in Lemma \ref{lemma:1} to detect deviations from $\mathcal{H}_0$. We do so by means of the empirical version (from an iid sample from \eqref{eq:modreg}) of the doubly-projected integrated regression function in \ref{lemma:1:7}, that is, the residual marked empirical process
\begin{align}
R_{n} \left(u, \gamma_{\mathcal{X}},\gamma_{\mathcal{Y}} \right) = \frac{1}{\sqrt{n}}  \sum_{i=1}^{n} \langle \hat{\mathcal{E}}_i, \gamma_{\mathcal{Y}} \rangle_{\mathbb{H}_2} \mathbbm{1}_{\left\lbrace \langle \mathcal{X}, \gamma_{\mathcal{X}} \rangle_{\mathbb{H}_1} \leq u \right\rbrace},
\label{eq_8}
\end{align}
with $u \in \mathbb{R},\,\gamma_{\mathcal{X}} \in \mathbb{S}_{\mathbb{H}_1},\,\gamma_{\mathcal{Y}} \in \mathbb{S}_{\mathbb{H}_2}$ and 
with residual marks $\langle \hat{\mathcal{E}}_i, \gamma_{\mathcal{Y}} \rangle_{\mathbb{H}_2} = \langle \mathcal{Y}_i - \langle \langle \mathcal{X}_i , \hat{\beta} \rangle \rangle, \gamma_{\mathcal{Y}} \rangle_{\mathbb{H}_2}$ and jumps $\langle \mathcal{X}_i, \gamma_{\mathcal{X}} \rangle_{\mathbb{H}_1}$, $i=1,\ldots,n$. To measure how close the empirical process \eqref{eq_8} is to zero, and following the ideas in \cite{Escanciano06} and \cite{GarciaPortuguesetal14}, we consider a Cram\'er--von Mises (CvM) norm on the space $\Pi = \mathbb{S}_{\mathbb{H}_2} \times \mathbb{S}_{\mathbb{H}_1} \times \mathbb{R} $, yielding what we term the Projected Cram\'er--von Mises (PCvM) statistic:
\begin{align}
\mathrm{PCvM}_n = &\; \int_{\Pi} \left[R_{n} \left(u,\gamma_{\mathcal{X}},\gamma_{\mathcal{Y}} \right) \right]^2 \,F_{n,\gamma_{\mathcal{X}}}(\mathrm{d}u)\,\omega_{\mathcal{X}}(\mathrm{d}\gamma_{\mathcal{X}})\,\omega_{\mathcal{Y}}(\mathrm{d}\gamma_{\mathcal{Y}}), \label{eq_9}
\end{align}
where $F_{n,\gamma_{\mathcal{X}}}$ is the empirical cumulative distribution function (ecdf) of $\lbrace \langle \mathcal{X}_i, \gamma_{\mathcal{X}} \rangle_{\mathbb{H}_1}\rbrace_{i=1}^n $, and $\omega_{\mathcal{X}} $ and $\omega_{\mathcal{Y}}$ are suitable measures on $\mathbb{S}_{\mathbb{H}_1}$ and $\mathbb{S}_{\mathbb{H}_2}$, respectively. As will be seen in Section \ref{sec:32}, a key advantage of the PCvM statistic with respect to other possible norms for \eqref{eq_8}, such as the Kolmogorov--Smirnov norm, is that it admits an explicit representation.\\

The infinite dimension of $\mathbb{S}_{\mathbb{H}_1}$ and $\mathbb{S}_{\mathbb{H}_2}$ makes the functional in \eqref{eq_9} unworkable. A way of circumventing this issue, motivated by Lemma \ref{lemma:2}, is to work with the finite-dimensional directions $\gamma^{(p)}_{\mathcal{X}}$ and $\gamma^{(q)}_{\mathcal{Y}}$ expressed on the bases $\{\Psi_j\}_{j=1}^p$ and $\{\Phi_k\}_{k=1}^q$, respectively. For the sake of simplicity, we assume that these bases are orthonormal from now on; see Remark \ref{rem:2} for non-orthogonal bases. Then, the $(p,q)$-truncated version of \eqref{eq_8} is
\begin{align*}
R_{n,p,q} \left(u, \gamma_{\mathcal{X}}^{(p)},\gamma_{\mathcal{Y}}^{(q)} \right) = \frac{1}{\sqrt{n}}
\sum_{i=1}^{n} \big\langle \hat{\mathcal{E}}_{i}^{(q)} , \gamma_{\mathcal{Y}}^{(q)} \big\rangle_{\mathbb{H}_2} \mathbbm{1}_{\left\lbrace \langle \mathcal{X}_{i}^{(p)}, \gamma_{\mathcal{X}}^{(p)}\rangle_{\mathbb{H}_1} \leq u \right\rbrace} = \frac{1}{\sqrt{n}}
\sum_{i=1}^{n} \hat{\mathbf{e}}_{i,q}' \mathbf{h}_q \mathbbm{1}_{\left\lbrace \mathbf{x}_{i,p}' \mathbf{g}_p \leq u \right\rbrace},
\end{align*}
where $u \in \mathbb{R},\,\mathbf{g}_p \in \mathbb{S}^{p-1},\,\mathbf{h}_q \in \mathbb{S}^{q-1}$ and $\hat{\mathbf{e}}_{i,q}'$ represents the $i$-th row of the $n\times q$ matrix of residual coefficients $\hat{\mathbf{E}}_{q}$, $\mathbf{g}_p$ and $\mathbf{h}_q$ are the coefficients of $\gamma_{\mathcal{X}}^{(p)}$ and $\gamma_{\mathcal{Y}}^{(q)}$, respectively, and $\mathbf{x}_{i,p}$ are the coefficients of $\mathcal{X}_i^{(p)}$. Therefore, the $(p,q)$-truncated version of \eqref{eq_9} is
\begin{align}
\mathrm{PCvM}_{n,p,q}
= &\; \int_{\Pi^{(p,q)}} \left[R_{n,p,q} \left(u, \gamma_{\mathcal{X}}^{(p)},\gamma_{\mathcal{Y}}^{(q)} \right) \right]^2  \,F_{n,\gamma_{\mathcal{X}}^{(p)}} (\mathrm{d}u) \,\omega_{\mathcal{X}} (\mathrm{d}\gamma_{\mathcal{X}}^{(p)}) \,\omega_{\mathcal{Y}}(\mathrm{d}\gamma_{\mathcal{Y}}^{(q)}),
\label{eq_5}
\end{align}
where $\Pi^{(p,q)} = \mathbb{S}_{\mathbb{H}_2,\{\Phi_k\}_{k=1}^\infty}^{q-1}\times \mathbb{S}_{\mathbb{H}_1,\{\Psi_j\}_{j=1}^\infty}^{p-1}\times \mathbb{R}$.

\subsection{Computable form of the statistic}
\label{sec:32}

The statistic in \eqref{eq_5} is now conveniently rewritten for its implementation. First, following \cite{Escanciano06} and \cite{GarciaPortuguesetal14}, let us assume that $\omega_{\mathcal{X}}$ and $\omega_{\mathcal{Y}}$ in \eqref{eq_5} represent uniform measures on $\mathbb{S}_{\mathbb{H}_1,\{\Psi_j\}_{j=1}^\infty}^{p-1}$ and $\mathbb{S}_{\mathbb{H}_2,\{\Phi_k\}_{k=1}^\infty}^{q-1}$, respectively. Second, recall that since both bases are orthonormal, from the transformation defined in \eqref{eq_1}, we have
\begin{align}
\mathrm{PCvM}_{n,p,q}  = &\;\int_{ \mathbb{S}^{q-1} \times \mathbb{S}^{p-1} \times \mathbb{R} } \left[R_{n,p,q} \left(u,\mathbf{g}_p, \mathbf{h}_q \right) \right]^2 \,F_{n,\mathbf{g}_p} (\mathrm{d}u) \,\mathrm{d}\mathbf{g}_p \,\mathrm{d}\mathbf{h}_q,\label{eq:pcvm2}
\end{align}
where $R_{n,p,q} \left(u,\mathbf{g}_p, \mathbf{h}_q \right)\equiv R_{n,p,q} \left(u, \gamma_{\mathcal{X}}^{(p)},\gamma_{\mathcal{Y}}^{(q)} \right)$. Using some simple algebra, we obtain
\begin{align}
\mathrm{PCvM}_{n,p,q}=&\;\int_{ \mathbb{S}^{q-1} \times \mathbb{S}^{p-1} \times \mathbb{R} } \frac{1}{n} \left[  \sum_{i=1}^{n} \hat{\mathbf{e}}_{i,q}' \mathbf{h}_q \mathbbm{1}_{\left\lbrace \mathbf{x}_{i,p}' \mathbf{g}_p \leq u \right\rbrace} \right]^2 \,F_{n,\mathbf{g}_p} (\mathrm{d}u) \,\mathrm{d}\mathbf{g}_p \,\mathrm{d}\mathbf{h}_q\nonumber\\
=&\; \frac{1}{n}  \sum_{i=1}^{n}  \sum_{j=1}^{n} \bigg[ \int_{ \mathbb{S}^{p-1} \times \mathbb{R} } \mathbbm{1}_{\left\lbrace \mathbf{x}_{i,p}' \mathbf{g}_p \leq u \right\rbrace}\mathbbm{1}_{\left\lbrace \mathbf{x}_{j,p}' \mathbf{g}_p \leq u \right\rbrace} \,F_{n,\mathbf{g}_p}(\mathrm{d}u) \,\mathrm{d} \mathbf{g}_p \bigg]\int_{ \mathbb{S}^{q-1} } (\hat{\mathbf{e}}_{i,q}' \mathbf{h}_q) (\hat{\mathbf{e}}_{j,q}' \mathbf{h}_q) \,\mathrm{d} \mathbf{h}_q \nonumber\\
=&\; \frac{1}{n^2}  \sum_{i=1}^{n} \bigg[ \sum_{j=1}^{n}  \sum_{r=1}^{n} \int_{ \mathbb{S}^{p-1}} \mathbbm{1}_{ \left\lbrace \left( \mathbf{x}_{i,p} - \mathbf{x}_{r,p} \right)' \mathbf{g}_p \leq 0,\, \left( \mathbf{x}_{j,p} - \mathbf{x}_{r,p} \right)' \mathbf{g}_p \leq 0 \right\rbrace} \,\mathrm{d} \mathbf{g}_p \bigg] E_{ij} \nonumber\\
=&\; \frac{1}{n^2}  \sum_{i=1}^{n}  \sum_{j=1}^{n}  \sum_{r=1}^{n} \left[ \int_{ S_{ijr}} \,\mathrm{d} \mathbf{g}_p \right]  E_{ij},\label{eq:pcvm1}
\end{align}
where we denote $S_{ijr} := \{ \textbf{z} \in \mathbb{S}^{p-1}:\pi/2 \leq \allowbreak\measuredangle \left( \mathbf{x}_{i,p} - \mathbf{x}_{r,p}, \textbf{z} \right) \leq 3\pi/2,\pi/2 \leq \measuredangle \left( \mathbf{x}_{j,p} - \mathbf{x}_{r,p}, \textbf{z}\right) \leq 3\pi/2 \}$ ($\measuredangle \left(\mathbf{x},\mathbf{y} \right)$ stands for the angle between $\mathbf{x},\mathbf{y} \in \mathbb{R}^{p}$) and $E_{ij}:=\int_{ \mathbb{S}^{q-1} } (\hat{\mathbf{e}}_{i,q}' \mathbf{h}_q) (\hat{\mathbf{e}}_{j,q}' \mathbf{h}_q) \,\mathrm{d} \mathbf{h}_q$.\\

The terms $\int_{S_{ijr}} \,\mathrm{d} \mathbf{g}_p =: A_{ijr}$ are the same as the ones given in \cite{GarciaPortuguesetal14} and they represent surface areas of particular spherical regions, that can either be the whole sphere $\mathbb{S}^{p-1}$ ($\mathbf{x}_{i,p}= \mathbf{x}_{j,p}=\mathbf{x}_{r,p}$), a hemisphere of $\mathbb{S}^{p-1}$ (if either $\mathbf{x}_{i,p}=\mathbf{x}_{j,p}$, $\mathbf{x}_{j,p}=\mathbf{x}_{r,p}$ or $\mathbf{x}_{i,p}=\mathbf{x}_{r,p}$), or a spherical wedge with solid angle
\begin{align}
\pi-\cos^{-1}\left( \frac{(\mathbf{x}_{i,p}-\mathbf{x}_{r,p})'(\mathbf{x}_{j,p}-\mathbf{x}_{r,p})}{\|\mathbf{x}_{i,p}-\mathbf{x}_{r,p}\| \cdot\|\mathbf{x}_{j,p}-\mathbf{x}_{r,p}\|} \right).
\label{spherical_wedge_eq}
\end{align}

Therefore, since the surface area of $\mathbb{S}^{p-1}$ is equal to  $2\pi^{p/2}/\Gamma \left(p/2\right)$, being $\Gamma \left( \cdot \right)$ the Gamma function, from \cite{Escanciano06} it follows that
\begin{align}
A_{ijr}= A_{ijr}^{(\measuredangle)}\frac{\pi^{p/2-1}}{\Gamma(p/2)},\quad 
A_{ijr}^{(\measuredangle)}:=\left\lbrace \begin{array} {ll} 2\pi, & \text{if }\mathbf{x}_{i,p} = \mathbf{x}_{j,p}=\mathbf{x}_{r,p},\\ \pi, & \text{if }\mathbf{x}_{i,p} \neq \mathbf{x}_{j,p}\text{ and }  \mathbf{x}_{i,p} = \mathbf{x}_{r,p} \text{ or } \mathbf{x}_{j,p}=\mathbf{x}_{r,p}, \\ \eqref{spherical_wedge_eq}, & \mathrm{otherwise.} \end{array}
\right. \label{eq:Aijr}
\end{align}

The term $E_{ij}$ can be dealt using the next auxiliary lemma regarding integration on the Euclidean sphere, yielding $E_{ij} = 2\pi^{q/2}/\left(q \Gamma \left(q/2 \right)\right) \hat{\mathbf{e}}_{i,q}' \hat{\mathbf{e}}_{j,q}$, for each $i,j=1,\ldots,n$.
\begin{lemma}\label{lemma:3}
	For any vectors $\mathbf{x},\mathbf{y} \in \mathbb{R}^q$, $\int_{\mathbb{S}^{q-1}} (\mathbf{x}'\boldsymbol{\omega}) (\mathbf{y}'\boldsymbol{\omega}) \,\mathrm{d} \boldsymbol{\omega} = 2 \pi^{q/2}/\left(q \Gamma \left(q/2\right)\right) \mathbf{x}'\mathbf{y}$.
\end{lemma}

Substituting these terms into \eqref{eq:pcvm1}, we get an easily computable form of the statistic:
\begin{align}
\mathrm{PCvM}_{n,p,q} = \frac{1}{n^2}  \sum_{i=1}^{n}  \sum_{j=1}^{n}  \sum_{r=1}^{n} A_{ijr} \frac{2\pi^{q/2}}{q \Gamma(q/2)} \hat{\mathbf{e}}_{i,q}' \hat{\mathbf{e}}_{j,q} = \frac{1}{n^2} \frac{2\pi^{p/2 + q/2 -1}}{q \Gamma(p/2) \Gamma(q/2)} {\rm Tr} \left[ \hat{\mathbf{E}}_{q}' \mathbf{A}_{\bullet} \hat{\mathbf{E}}_{q} \right], \label{eq:pcvm3}
\end{align}
where ${\rm Tr}(\cdot)$ denotes the trace operator and the elements of the symmetric matrix $\mathbf{A}_{\bullet}$ are defined as $ \left(\mathbf{A}_{\bullet} \right)_{ij} :=  \sum_{r=1}^{n} A_{ijr}$, for $i,j=1,\ldots,n$.

\begin{remark}[Generalization of the GoF test statistic for the FLMSR]
	If $\mathbb{H}_2=\mathbb{R}$, identifiable with the subspace of $L^2([c,d])$ of constant functions, the FLMSR arises as a particular case of the FLMFR. This is reflected in the statistic \eqref{eq:pcvm3} which, if $q=1$, yields the PCvM statistic for the FLMSR given in \cite{GarciaPortuguesetal14} as a particular case.
\end{remark}

\begin{remark}[Alternative interpretation of $\mathrm{PCvM}_{n,p,q}$] 
	\label{rem:3}
	The statistic \eqref{eq:pcvm3} can be written as
	\begin{align*}
	\mathrm{PCvM}_{n,p,q}&= \frac{1}{n^2} \frac{2\pi^{p/2 + q/2 -1}}{q \Gamma(p/2) \Gamma(q/2)}  \sum_{k=1}^{q} \left\| \left(\hat{e}_{1,k}, \ldots, \hat{e}_{n,k} \right)  \right\|_{\mathbf{A}_{\bullet}}\!,
	\end{align*}
	where $\hat{\mathcal{E}}_{i}^{(q)} = \sum_{k=1}^{q} \hat{e}_{i,k} \Phi_{k}$, $i=1,\ldots,n$, and $\left\| \mathbf{v} \right\|_{\mathbf{A}_{\bullet}}:=(\mathbf{v}'\mathbf{A}_\bullet\mathbf{v})^{1/2}$ is a norm (see Lemma \ref{lem:4}) in $\mathbb{R}^{n}$ induced by the symmetric matrix $\mathbf{A}_{\bullet}$. Therefore, the statistic is a sum, across the $q$ dimensions of the truncated response, of the $\mathbf{A}_\bullet$-weighted norms of the coefficients of the functional errors on $\{\Phi_k\}_{k=1}^q$. If this basis is non-orthonormal, then a similar interpretation can be obtained (see Remark~\ref{rem:2}).
\end{remark}

Observe that $\|\cdot\|_{\mathbf{A}_\bullet}$ is trivially a semi-norm: since $\mathrm{PCvM}_{n,p,q}$ is non-negative, then $\mathbf{A}_\bullet$ must be positive semi-definite. That $\mathbf{A}_\bullet$ is actually a norm follows from the next lemma.

\begin{lemma}\label{lem:4}
	Assume that the functional sample $\{\mathcal{X}_i\}_{i=1}^n$ has pairwise \emph{distinct} coefficients $\{\mathbf{x}_{i,p}\}_{i=1}^n$ on an arbitrary $p$-truncated basis $\lbrace \Psi_{j}\rbrace_{j=1}^p$ of $\mathbb{H}_1$. Then, for any sample size $n\geq1$, the $n\times n$ matrix $\mathbf{A}_\bullet$ is positive definite.
\end{lemma}

\begin{remark}[Statistic for general functional bases]
	\label{rem:2}
	The statistic in \eqref{eq:pcvm2} can be expressed in terms of non-orthogonal functional bases as follows:
	\begin{align*}
	|\mathbf{P}_p | \left| \mathbf{Q}_q \right| \mathrm{PCvM}_{n,p,q}
	=&\;\int_{ \mathbb{S}^{q-1} \times \mathbb{S}^{p-1} \times \mathbb{R} } \left[R_{n,p,q} \left(u,\mathbf{P}_p^{-1}\mathbf{g}_p, \mathbf{Q}_q^{-1}\mathbf{h}_q \right) \right]^2\,F_{n,\mathbf{P}_p^{-1}\mathbf{g}_p} (\mathrm{d}u) \,\mathrm{d}\mathbf{g}_p \,\mathrm{d}\mathbf{h}_q\\
	=&\; \int_{ \mathbb{S}^{q-1} \times \mathbb{S}^{p-1} \times \mathbb{R} } \frac{1}{n} \left[  \sum_{i=1}^{n} \hat{\mathbf{e}}_{i,q}' \mathbf{Q}_q' \mathbf{h}_q \mathbbm{1}_{\left\lbrace \mathbf{x}_{i,p}' \mathbf{P}_p' \mathbf{g}_p \leq u \right\rbrace} \right]^2 \,F_{n,\mathbf{P}_p^{-1}\mathbf{g}_p} (\mathrm{d}u) \,\mathrm{d}\mathbf{g}_p \,\mathrm{d}\mathbf{h}_q,
	\end{align*}
	where $\boldsymbol{\Phi}=\mathbf{Q}_q'\mathbf{Q}_q$ is the Cholesky decomposition of $\boldsymbol{\Phi}$ and the second equality stems from $\langle \mathcal{X}_{i}^{(p)}, \gamma_{\mathcal{X}}^{(p)} \rangle_{\mathbb{H}_1} \allowbreak= \mathbf{x}_{i,p}' \boldsymbol{\Psi}\mathbf{g}_p$ and $\langle \hat{\mathcal{E}}_{i}^{(q)}, \gamma_{\mathcal{Y}}^{(q)} \rangle_{\mathbb{H}_2}  = \hat{\mathbf{e}}_{i,q}' \boldsymbol{\Phi}\mathbf{h}_q$. Then, following the developments preceding \eqref{eq:pcvm3}, it can be shown that
	\begin{align}
	\mathrm{PCvM}_{n,p,q} = \frac{1}{n^2} \frac{2\pi^{p/2 + q/2 -1}}{|\mathbf{P}_p | \left| \mathbf{Q}_q \right|q \Gamma(p/2) \Gamma(q/2)} {\rm Tr} \left[ (\hat{\mathbf{E}}_{q} \mathbf{Q}_q )'  \mathbf{A}_{\bullet}
	(\hat{\mathbf{E}}_{q} \mathbf{Q}_q ) \right], \label{gen_stat}
	\end{align}
	where $\mathbf{A}_{\bullet}$ is based on the coefficients of $\mathcal{X}_1^{(p)},\ldots,\mathcal{X}_n^{(p)}$ on the non-orthonormal basis $\{\Psi_j\}_{j=1}^p$.
\end{remark}

Despite the general derivation of the PCvM statistic, we will focus on its application for the data-driven FPC bases $\{\hat\Psi_j\}_{j=1}^n$ and $\{\hat\Phi_k\}_{k=1}^n$.

\subsection{Testing in practice and bootstrap resampling}
\label{sec:34}

We calibrate the null distribution of the statistic $\mathrm{PCvM}_{n,p,q}$ in \eqref{gen_stat} by a wild bootstrap on the residuals. This methodology is consistent in the finite dimensional case, as shown by \cite{Stuteetal98}, and well-adapted for heteroscedastic scenarios.\\

The bootstrap resampling is detailed within the next algorithm. It describes how to perform our GoF test proposal in practice using FPCR-L1S, as this estimator combines the performance of FPCR-L1 and the computational expediency of FPCR. Adaptations to other estimators described in Section \ref{sec:23} are straightforward (but see Remark \ref{rem:4} below).

\begin{algo}[Testing in practice]\label{algo}
	Let $\lbrace \left(\mathcal{X}_i,\mathcal{Y}_i \right)\rbrace_{i=1}^n$ be an iid sample. The GoF test for the FLMFR proceeds as follows:
	\begin{enumerate}[label=\arabic*., ref=\textit{\arabic*}]
		\item Center the sample and compute the FPC of $\lbrace \mathcal{X}_i\rbrace_{i=1}^n$ and $\lbrace \mathcal{Y}_i \rbrace_{i=1}^n$.\label{algo:1}
		
		\item Select $p$ and $q$ as the minimum number of FPC required for attaining a certain proportion of EV (e.g., such that $\mathrm{EV}_p=\mathrm{EV}_q=0.99$).\label{algo:2}
		
		\item Compute the coefficients (scores) of $\lbrace \mathcal{X}_i\rbrace_{i=1}^n$ and $\lbrace \mathcal{Y}_i \rbrace_{i=1}^n$ on the $p$- and $q$-truncated FPC bases, resulting the matrices $\mathbf{X}_p$ and $\mathbf{Y}_q$.\label{algo:3}
		
		\item Compute the FPCR-L1S estimator $\hat{\mathbf{B}}^{(\lambda),\mathrm{C}}_{\tilde{p},q}$ of $\beta$ as described in Section \ref{sec:23}. This automatically selects a subset of $\tilde{p}$ out of $p$ FPC coefficients, depending on $\lambda$.\label{algo:4}
		
		\item Obtain the residuals
		$
		\hat{\mathbf{e}}_{i,q} = \mathbf{Y}_{i,q} - \mathbf{X}_{i,p} \hat{\mathbf{B}}^{(\lambda),\mathrm{C}}_{\tilde{p},q}$, $ i=1,\ldots,n,
		$
		and compute with them the statistic   $\mathrm{PCvM}_{n,\tilde{p},q}$ in \eqref{eq:pcvm3}.\label{algo:5}
		
		\item Perform the bootstrap resampling. For $b=1,\ldots,B$:\label{algo:6}
		\begin{enumerate}[label=\roman*., ref=\textit{\roman*}]
			
			\item Simulate independent zero-mean and unit-variance random variables $\lbrace V_{i}^{\ast b}\rbrace_{i=1}^{n}$. For example, sample $V^{\ast b}$ such that
			$\mathbb{P}\left[ V^{\ast b}=(1\mp\sqrt{5})/2\right]=(5\pm\sqrt{5})/10$.\label{algo:i}
			
			\item Set the bootstrap errors as $\mathbf{e}_{i,q}^{\ast b}:=\hat{\mathbf{e}}_{i,q} V_{i}^{\ast b}$, $i=1,\ldots,n$.\label{algo:ii}
			
			\item Set the \emph{uncentered} bootstrapped responses $\mathbf{Y}_{i,q}^{\ast b,u}:=\mathbf{X}_{i,\tilde{p}}\hat{\mathbf{B}}^{(\lambda),\mathrm{C}}_{\tilde{p},q}+\mathbf{e}_{i,q}^{\ast b}$, and center them to imitate the original FPC scores: $\mathbf{Y}_{i,q}^{\ast b}:=\mathbf{Y}_{i,q}^{\ast b,u}-\overline{\mathbf{Y}_{q}^{\ast b,u}}$, $i=1,\ldots,n$.\label{algo:iii}
			
			\item From the bootstrap sample $\{(\mathbf{X}_{i,\tilde{p}},\mathbf{Y}_{i,q}^{\ast b})\}_{i=1}^n$, compute the estimator $\hat{\mathbf{B}}_{\tilde{p},q}^{\ast b}$ of $\hat{\mathbf{B}}^{(\lambda),\mathrm{C}}_{\tilde{p},q}$.\label{algo:iv}
			
			\item Obtain the bootstrap residuals
			$\hat{\mathbf{e}}_{i,q}^{\ast b}=\mathbf{Y}_{i,q}^{\ast b}-\mathbf{X}_{i,\tilde{p}}\hat{\mathbf{B}}_{\tilde{p},q}^{\ast b}$, $ i=1,\ldots,n$,
			and compute with them the bootstrapped statistic $\mathrm{PCvM}^{\ast b}_{n,\tilde{p},q}$ from \eqref{eq:pcvm3}.\label{algo:v}
		\end{enumerate}
		
		\item Estimate the $p$-value by Monte Carlo as usual by $\#\lbrace \mathrm{PCvM}_{n,\tilde{p},q} \leq \mathrm{PCvM}_{n,\tilde{p},q}^{\ast b} \rbrace / B$.
	\end{enumerate}
\end{algo}

\begin{remark}[Computational tricks]\label{rem:4}
	Since $\mathbf{A}_{\bullet}$ depends exclusively on the covariate sample, it only needs to be computed once in the testing procedure. In addition, as the wild bootstrap only affects the response, steps \ref{algo:iv}--\ref{algo:v} can be efficiently implemented using the hat matrix \eqref{hat_matrix_FPCR-L2-C}, avoiding costly refittings on each bootstrap iteration. Indeed, $\hat{\mathbf{E}}_q^{*b}=\mathbf{Y}_{q}^{*b}-\hat{\mathbf{Y}}_{q}^{*b}=\big(\mathbf{I}_q-\mathbf{H}_\mathrm{C}^{(\lambda)}\big)\mathbf{Y}_{q}^{*b}$, $ \hat{\mathbf{Y}}_{q}^{*b}=\mathbf{X}_{\tilde{p}}\hat{\mathbf{B}}_{\tilde{p},q}^{\ast b}$. The same comment holds for FPCR-L2 and FPCR by virtue of $\mathbf{H}^{(\lambda)}$ (in that case, $\tilde{p}=p$), although not for FPCR-L1 due to its lack of an explicit hat matrix. The GoF test using FPCR-L1 thus requires $B+1$ LASSO fits.
\end{remark}

\begin{remark}[Scores versus functional resampling]
	The above wild bootstrap performs the resampling on the \emph{scores} of the residuals in the $q$-truncated FPC basis $\{\hat{\Phi}_k\}_{k=1}^q$, as from step \ref{algo:4} onwards there is no further mention to the functional nature of the sample. This view could be achieved with extra notation, as the bootstrap errors in step \ref{algo:ii} can be written as \begin{align*}
	\mathcal{E}_i^{(q)\ast b}:=\sum_{k=1}^q(e_{i,k}V_i^{\ast b})\hat\Phi_k, \quad \hat{\mathcal{E}}_i^{(q)}=\sum_{k=1}^qe_{i,k}\hat\Phi_k.
	\end{align*}
	This exposes a subtle point: why not bootstrapping the \emph{functional} residuals $\hat{\mathcal{E}}_i=\mathcal{Y}_i-\hat{\mathcal{Y}}^{(q)}_i$ as $\mathcal{E}^{\ast b}_i:=\hat{\mathcal{E}}_i V_i^{\ast b}$? This would allow to obtain truly functional bootstrap responses $\{\mathcal{Y}_i^{\ast b}\}_{i=1}^n$, yet at expenses of the overhead of recomputing their FPC for each bootstrap replicate. In our experiments, this latter approach did not provide a significant improvement on the calibration of the test over the scores resampling, hence it was discarded in favor of the latter.
\end{remark}

\begin{remark}[Selection of the penalty parameter]\label{rem:6}
	A possible data-driven selection for $\lambda$ in step \ref{algo:4} is $\hat\lambda_{\mathrm{CV}}$. However, we found by simulations that the so-called \emph{one standard error rule} $\hat\lambda_{1\mathrm{SE}}$ (see, e.g., \cite{Friedman2010}) improved the stability of the calibration of $\mathrm{PCvM}_{n,\tilde{p},q}$ under $\mathcal{H}_0$. This is coherent with the folklore in smoothing-based GoF tests, where the optimal smoothing parameter for estimating the regression function $m$ is often \emph{not} the most appropriate for conducting the test; instead, an oversmoothed estimate of $m$ (that biases the estimation in exchange for a variance reduction, precisely as $\hat\lambda_{1\mathrm{SE}}$ does) is desirable for a better calibration of the statistic.
\end{remark}

\begin{remark}[FPCR-L1 variable selection]\label{rem:7}
	LASSO is a consistent variable selector if the predictors are \emph{uncorrelated} \citep{Zhao2006}. Hence, this result supports the adequateness for combining FPCR (instead of using non-orthogonal bases) with LASSO variable-selection. It also supports ignoring in Algorithm \ref{algo} the bootstrapping of the variable selection uncertainty. Indeed, motivated by a comment of one referee, we ran a small simulation study with a modified version of Algorithm \ref{algo} that incorporated in Step \ref{algo:6} a bootstrap variable selection using $\hat{\lambda}_{\mathrm{1SE}}^{*b}$, obtaining very similar powers to the analogs of Tables \ref{tab:res_comp_S1}--\ref{tab:res_comp_S3} for FPCR-L1S ($\hat{\lambda}_{\mathrm{1SE}}$). Clearly, this modification increases the computational requirements by orders of magnitude, which is impractical. In this simulation it was also evidenced that variable selection based on $\hat{\lambda}_{\mathrm{CV}}$ seems to be inconsistent (which may be explained by \cite{Shao1993}'s result), while variable selection based on $\hat{\lambda}_{\mathrm{1SE}}$ seems to be behave consistently.
\end{remark}

So far we have only discussed the GoF test for the FLMFR. However, simple adaptations allow to test also the simple hypothesis $\mathcal{H}_0:m(\cdot) = \langle \langle \cdot, \beta_0 \rangle \rangle$, where $\beta_0\in\mathbb{H}_1\otimes\mathbb{H}_2$ now is specified. Algorithm \ref{algo} can be straightforwardly adapted. First, replace step \ref{algo:4} by
\textit{\begin{enumerate}[label=\arabic*'., ref=\textit{\arabic*'}]
		\setcounter{enumi}{3}
		\item Compute $\tilde{p}$ as in step \ref{algo:4}. Obtain $\mathbf{B}^0_{\tilde{p},q}=(b_{ij}^0)_{ij}$, the $\tilde{p}\times q$ matrix of $\beta_0$ FPC coefficients.
\end{enumerate}}
Then, the bootstrap procedure is subsequently adjusted by simply ignoring the estimation steps, that is, by replacing both $\hat{\mathbf{B}}^{(\lambda),\mathrm{C}}_{\tilde{p},q}$ and $\hat{\mathbf{B}}_{\tilde{p},q}^{\ast b}$ by $\mathbf{B}^0_{\tilde{p},q}$.\\

Algorithm \ref{algo} and its variants (simple hypothesis; FPCR, FPCR-L2, and FPCR-L1 estimators; functional residual resampling) are implemented in the companion R package \texttt{goffda} \citep{Garcia-Portugues:goffda}. The critical parts of the test, such as the computation of the $\mathbf{A}_\bullet$ matrix and the computation of the PCvM statistic (whose complexity is $\mathcal{O} \left(q(n^3 - n^2)/2 \right)$), are implemented in C++ for the sake of efficiency.

\section{Simulation study}
\label{sec:4}

The finite sample behaviour of the PCvM test is now illustrated via a comparative study with the available significance tests (Section \ref{sec:41}) and a  simulation study for the composite hypothesis (Section \ref{sec:42}). We employed the scenarios already described in Table \ref{tab:scenarios} and used the following common settings: discretization of functional samples in $101$ equispaced grid points along the domains, sample sizes $n=50, 100, 250$, $B=1,000$, and $1,000$ Monte Carlo replicates. The PCvM test was run using Algorithm \ref{algo} with $\mathrm{EV}_p = \mathrm{EV}_q = 0.99$.\\

The PCvM test was computed using both FPCR and FPCR-L1S, for showing how the overfitting inherent to the former may affect the GoF test. In Section \ref{sec:42}, FPCR-L1S is employed with both $\hat\lambda_{\mathrm{CV}}$ and $\hat\lambda_{1\mathrm{SE}}$ for the purpose of illustrating the discussion in Remark \ref{rem:6}. When testing for significance, the conclusions reached with both penalty parameters were similar (since an estimator of $\beta$ is not required), so the results are only reported for $\hat\lambda_{1\mathrm{SE}}$. The search for $\hat\lambda_{\mathrm{CV}}$ and $\hat\lambda_{1\mathrm{SE}}$ was done among a sequence in $[10^{-3},10^{2}]$ and, if the minimizer of the objective function was found at its extremes, the interval was expanded. 

\subsection{Simple hypothesis}
\label{sec:41}

We compare in this section the significance tests by \cite{Kokoszkaetal08}, \cite{Patileaetal16}, and \cite{Leeetal2020} (henceforth abbreviated as KMSZ, PSS, and LZS, respectively) with our PCvM test for the no effects hypothesis 
\begin{align*}
\mathcal{H}_{0,\mathrm{NE}}:m(\cdot) = \langle \langle \cdot, \beta_0 \rangle \rangle, \quad \beta_0 \equiv 0.
\end{align*}

Both the KMSZ and PSS tests are based on the FPC of the predictor and response, that are truncated such that $\mathrm{EV}_p = \mathrm{EV}_q = 0.99$. The KMSZ statistic is asymptotically $\chi_{pq}^{2}$ distributed under $\mathcal{H}_{0,\mathrm{NE}}$, this being the distribution employed to calibrate the test. We ran the PSS test as implemented in the \texttt{fdapss} \citep{Patileaetal16:fdapss} package, with a grid of $50$ points for each one-dimensional optimization and the bandwidth chosen as $h = n^{-2/9}$, as suggested in \cite{Patileaetal16}. A bug in \texttt{pss.test} when $p = 1$ invalidated up to $8.5\%$ of the Monte Carlo replicates, depending on the\nolinebreak[4] scenario.\\

The LZS test estimates the functional martingale difference divergence that characterizes the conditional mean dependence of $\mathcal{X}$ and $\mathcal{Y}$. Hence, unlike the previous competitors and our approach, it does not require from an FPC-based dimension reduction. 

\begin{table}[h!]
	\iffigstabs
	\centering
	\footnotesize
	\begin{tabular}{ccl}
		\toprule
		Notation & Model & $\left(\delta_1, \delta_2, \delta_3 \right)$ \\
		\midrule
		$\mathcal{H}_{0,\mathrm{NE}}$ (no effects) & $\mathcal{Y}(t) = \mathcal{E}(t)$ & None \\
		\midrule
		\multirow{3}{*}{$\mathcal{H}_{1,\mathrm{FR}}^{h}$ (FLMFR)} & \multirow{3}{*}{$\mathcal{Y}(t) =  \delta_h \langle \langle \mathcal{X}, \beta\rangle \rangle + \mathcal{E}(t) $} & {S1}: $\left(0.035, 0.08, 0.15\right)$  \\
		& & {S2}: $\left(0.01, 0.02, 0.03\right)$ \\
		& & {S3}: $\left(1, 1.3, 1.6\right)$ \\ 
		\midrule
		\multirow{3}{*}{$\mathcal{H}_{1,\mathrm{C}}^{h}$ (FLCFR)} & \multirow{3}{*}{$\mathcal{Y}(t) =   \delta_h  \widetilde{\beta}_j(t) \mathcal{X}(t)  + \mathcal{E}(t),\; j = 1,2,3$} & S1: $\left(0.025, 0.05, 0.15\right)$   \\
		& & S2: $\left(0.2, 0.6, 1\right)$   \\
		& & S3: $\left(0.01, 0.025, 0.05\right)$   \\
		\midrule  	
		$\mathcal{H}_{1,\mathrm{NLQ}}^{h}$ & $\mathcal{Y}(t) =   \delta_h \Delta \left( \mathcal{X} \right)(t) + \mathcal{E}(t)$ & \multirow{4}{4cm}{{S1}: $\left(0.025, 0.075, 0.15\right)$ \linebreak {S2}: $\left(0.02, 0.04, 0.1\right)$ \linebreak {S3}: $\left(0.2, 0.35, 0.55\right)$}    \\ 
		(non linear, quadratic) & $\Delta \left( \mathcal{X} \right)(t) =\mathcal{X}^2 \big(a + (t - c) \frac{b - a}{d - c}\big) - 1$ &  \\  
		\cmidrule{1-2}
		$\mathcal{H}_{1,\mathrm{NLT}}^{h}$ & $\mathcal{Y}(t) = \delta_h \Delta \left( \mathcal{X} \right)(t) + \mathcal{E}(t)$ &    \\
		(non linear, trigonometric) & $\Delta \left( \mathcal{X} \right)(t) = \left( \sin (2 \pi t)  - \cos (2 \pi t) \right) \left\| \mathcal{X} \right\|_{\mathbb{H}_1}^{2}$ &    \\
		\bottomrule
	\end{tabular}
	\caption{\small{Summary of null and alternative hypotheses. Concurrent models are given by functions $\widetilde{\beta}_1(t) = \sqrt{\left| \sin(\pi t)  - \cos(\pi t) \right|}$ (S1),  $\widetilde{\beta}_2(t) = \log \left(t - a + 0.5\right)$ (S2), and $\widetilde{\beta}_3(t) = \left(t-0.5 \right)^3$ (S3). \label{tab:hyp_simple}}}
	\fi
\end{table}

We assume here that $\mathbb{H}_1 =\mathbb{H}_2 = L^2 \left([0,1]\right)$. As reflected in Table \ref{tab:hyp_simple}, four  kind of deviations from $\mathcal{H}_{0,\mathrm{NE}}$ were generated: FLMFR, concurrent model (degenerated FLMFR, denoted as FLCFR), and two nonlinear alternatives. The empirical rejection rates are given in Tables \ref{tab:res_simple_S1}--\ref{tab:res_simple_S3}. They contain only the results of the FPCR-based PCvM test since the FPCR-L1S version gave almost identical rejection rates. Their analysis reveals the following insights:
\begin{itemize}
	
	\item Regarding the calibration, the PCvM and LZS tests are the only without repeated miscalibrations in any scenario: an over-rejection happens in S2 ($\mathcal{H}_{0,\mathrm{NE}}$ in Table \ref{tab:res_simple_S2}) for the PSS test, while the KMSZ test has difficulties in S2 and S3 ($\mathcal{H}_{0,\mathrm{NE}}$ in Tables \ref{tab:res_simple_S2}--\ref{tab:res_simple_S3}).
	
	\item Concerning linear alternatives (FLMFR and concurrent),  
	the LZS and KMSZ tests seem to be the most powerful in S1, but the KMSZ test is notably the most powerful approach in S2 under the FLMFR alternative, an outcome somehow expected given the test nature. However, the KMSZ test may fail under linear alternatives for sparse scenarios ($\mathcal{H}_{1,\mathrm{FR}}$ and $\mathcal{H}_{1,\mathrm{C}}$ in Table \ref{tab:res_simple_S3}), providing empirical powers smaller than the nominal level. The behaviour is worse for the LZS test, even under larger deviations from $\mathcal{H}_{0,\mathrm{NE}}$. A possible explanation is that the noise introduced with the null FPCs is not removed due to the lack of dimension reduction in the test. With respect to the comparison of the PCvM and PSS tests, the former is more powerful than the latter under concurrent models ($\mathcal{H}_{1,\mathrm{C}}$ in Tables \ref{tab:res_simple_S1}--\ref{tab:res_simple_S3}) in all scenarios and for all sample sizes. In the case of FLMFR alternatives, this is also the case (unless for minor exceptions) for S1 and S2 ($\mathcal{H}_{1,\mathrm{FR}}$ in Tables \ref{tab:res_simple_S1}--\ref{tab:res_simple_S2}). In S3, the PSS test attains perfect empirical power, even for $n=50$ and the smallest deviation from the null hypothesis, manifesting a sharp difference with respect to its behavior for S2 (almost blind for linear alternatives). 
	
	\item Concerning nonlinear alternatives, as expected, KMSZ exhibits a poor performance detecting them, except for S2 under $\mathcal{H}_{1,\mathrm{NLQ}}$. The PSS, LZS, and PCvM tests correctly detect all the nonlinear alternatives, the former being on overall more powerful in S3, the second one in S1 and S2.
	
\end{itemize}

\begin{table}[h!]
	\iffigstabs
	\setlength{\tabcolsep}{5pt}
	\centering
	\footnotesize
	\begin{tabular}{c|ccc|ccc|ccc|ccc}
		\addlinespace[-\aboverulesep] 
		\cmidrule[\heavyrulewidth]{2-13}
		\multicolumn{1}{c}{} & \multicolumn{3}{c}{KMSZ} & \multicolumn{3}{c}{PSS} & \multicolumn{3}{c}{LZS} & \multicolumn{3}{c}{$\mathrm{PCvM}$} \\	
		\midrule
		$n$ & $50$ & $100$ & $250$ & $50$ & $100$ & $250$ & $50$ & $100$ & $250$ & $50$ & $100$ & $250$ \\
		\midrule
		$\mathcal{H}_{0,\mathrm{NE}}$ & $\boldsymbol{0.053}$ & $\boldsymbol{0.055}$ & $\boldsymbol{0.057}$ & $\boldsymbol{0.049}$ & $\boldsymbol{0.047}$  & $\boldsymbol{0.048}$ & $\boldsymbol{0.048}$ & $\boldsymbol{0.044}$ & $\boldsymbol{0.049}$  & $\boldsymbol{0.042}$ & 0.034 & $\boldsymbol{0.050}$ \\  
		\midrule
		$\mathcal{H}_{1,\mathrm{FR}}^{1}$ & 0.083 & $\boldsymbol{0.178}$ & $\boldsymbol{0.495}$ & 0.067 & 0.074 & 0.160 & 0.339 & 0.567 & 0.943 & $\boldsymbol{0.087}$ & 0.128 & 0.282 \\
		$\mathcal{H}_{1,\mathrm{FR}}^{2}$ & $\boldsymbol{0.384}$ & $\boldsymbol{0.836}$ & $\boldsymbol{1.000}$ & 0.177 & 0.316 & 0.718 & 0.916 & 0.998 & 1.000 & 0.292 & 0.516 & 0.923 \\
		$\mathcal{H}_{1,\mathrm{FR}}^{3}$ & $\boldsymbol{0.955}$ & $\boldsymbol{1.000}$ & $\boldsymbol{1.000}$ & 0.551 & 0.885 & 0.997 & 1.000 & 1.000 & 1.000 & 0.718 & 0.973 & $\boldsymbol{1.000}$ \\   
		\midrule
		$\mathcal{H}_{1,\mathrm{C}}^{1}$ & $\boldsymbol{0.146}$ & $\boldsymbol{0.378}$ & $\boldsymbol{0.890}$ & 0.066 & 0.069 & 0.152 & 0.339 & 0.576 & 0.946 & 0.082 & 0.121 & 0.272 \\
		$\mathcal{H}_{1,\mathrm{C}}^{2}$ & $\boldsymbol{0.527}$ & $\boldsymbol{0.936}$ & $\boldsymbol{1.000}$ & 0.113 & 0.195 & 0.472 & 0.822 & 0.989 & 1.000 & 0.171 & 0.340 & 0.778 \\  
		$\mathcal{H}_{1,\mathrm{C}}^{3}$ & $\boldsymbol{0.976}$ & $\boldsymbol{1.000}$ & $\boldsymbol{1.000}$ & 0.708 & 0.969 & $\boldsymbol{1.000}$ & 1.000 & 1.000 & 1.000 & 0.511 & 0.864 & $\boldsymbol{1.000}$ \\
		\midrule
		$\mathcal{H}_{1,\mathrm{NLQ}}^{1}$ & 0.050 & $\boldsymbol{0.065}$ & 0.070 & 0.052 & 0.061 & $\boldsymbol{0.116}$ & 0.113 & 0.218 & 0.677 & $\boldsymbol{0.053}$ & 0.045 & 0.074 \\
		$\mathcal{H}_{1,\mathrm{NLQ}}^{2}$ & 0.125 & 0.171 & 0.168 & $\boldsymbol{0.143}$ & $\boldsymbol{0.362}$ & $\boldsymbol{0.876}$ & 0.581 & 0.970 & 1.000 & 0.086 & 0.171 & 0.686 \\
		$\mathcal{H}_{1,\mathrm{NLQ}}^{3}$ & 0.246 & 0.274 & 0.255 & $\boldsymbol{0.553}$ & $\boldsymbol{0.959}$ & $\boldsymbol{1.000}$ & 0.876 & 1.000 & 1.000 & 0.233 & 0.721 & $\boldsymbol{1.000}$ \\		
		\midrule
		$\mathcal{H}_{1,\mathrm{NLT}}^{1}$ & $\boldsymbol{0.100}$ & $\boldsymbol{0.135}$ & $\boldsymbol{0.129}$ & 0.050 & 0.050 & 0.059 & 0.093 & 0.133 & 0.502 & 0.047 & 0.039 & 0.064 \\
		$\mathcal{H}_{1,\mathrm{NLT}}^{2}$ & $\boldsymbol{0.194}$ & $\boldsymbol{0.217}$ & 0.196 & 0.068 & 0.132 & $\boldsymbol{0.791}$ & 0.632 & 0.987 & 1.000 & 0.080 & 0.107 & 0.483 \\
		$\mathcal{H}_{1,\mathrm{NLT}}^{3}$ & 0.217 & 0.237 & 0.216 &  $\boldsymbol{0.446}$ & $\boldsymbol{0.949}$ & $\boldsymbol{1.000}$ & 0.932 & 1.000 & 1.000 & 0.245 & 0.743 & $\boldsymbol{1.000}$ \\
		\bottomrule
	\end{tabular}
	\caption{\small{Scenario S1. Empirical rejection rates for the KMSZ, PSS, LZS, and PCvM tests for $n=50, 100, 250$ and the deviations in Table \ref{tab:hyp_simple}. Under $\mathcal{H}_{0,\mathrm{NE}}$, the rejection rates are boldfaced if they lie in the $95\%$-confidence interval of the nominal level, $0.05$. Under $\mathcal{H}_{1}$, boldfaces denote the empirical powers that are not significantly smaller than the largest, for each deviation and sample size, according to a $95\%$-confidence paired $t$-test. \label{tab:res_simple_S1}}}
	\fi
\end{table}

\begin{table}[h!]
	\iffigstabs
	\setlength{\tabcolsep}{5pt}
	\centering
	\footnotesize
	\begin{tabular}{c|ccc|ccc|ccc|ccc}
		\addlinespace[-\aboverulesep]
		\cmidrule[\heavyrulewidth]{2-13}
		\multicolumn{1}{c}{} & \multicolumn{3}{c}{KMSZ} & \multicolumn{3}{c}{PSS} & \multicolumn{3}{c}{LZS} & \multicolumn{3}{c}{$\mathrm{PCvM}$} \\	
		\midrule
		$n$ & $50$ & $100$ & $250$ & $50$ & $100$ & $250$ & $50$ & $100$ & $250$ & $50$ & $100$ & $250$ \\
		\midrule
		$\mathcal{H}_{0,\mathrm{NE}}$ & 0.006 & 0.033 & $\boldsymbol{0.043}$ & 0.093 & 0.070 & 0.068 & $\boldsymbol{0.054}$ & $\boldsymbol{0.048}$ & $\boldsymbol{0.047}$ & 0.030 & $\boldsymbol{0.036}$ & $\boldsymbol{0.045}$ \\
		\midrule
		$\mathcal{H}_{1,\mathrm{FR}}^{1}$ & 0.025 & $\boldsymbol{0.201}$ & $\boldsymbol{0.932}$ & $\boldsymbol{0.091}$ & 0.078 & 0.064 & 0.064 & 0.064 & 0.131 & 0.036 & 0.056 & 0.107 \\
		$\mathcal{H}_{1,\mathrm{FR}}^{2}$ & 0.058 & $\boldsymbol{0.521}$ & $\boldsymbol{1.000}$ & $\boldsymbol{0.094}$ & 0.078 & 0.057 & 0.125 & 0.259 & 0.952 & 0.065  & 0.168 & 0.900 \\
		$\mathcal{H}_{1,\mathrm{FR}}^{3}$ & 0.083 & 0.657 & $\boldsymbol{1.000}$ & 0.095 & 0.075 & 0.058 & 0.342 & 0.846 & 1.000  & $\boldsymbol{0.180}$ & \textbf{0.729} & $\boldsymbol{1.000}$ \\ 
		\midrule
		$\mathcal{H}_{1,\mathrm{C}}^{1}$ & 0.023 & $\boldsymbol{0.112}$ & $\boldsymbol{0.567}$ & $\boldsymbol{0.087}$ & 0.085 & 0.059 & 0.868 & 0.999 & 1.000 & 0.032 & 0.056 & 0.104 \\
		$\mathcal{H}_{1,\mathrm{C}}^{2}$ & $\boldsymbol{0.120}$ & $\boldsymbol{0.874}$ & $\boldsymbol{1.000}$ & 0.082 & 0.092 & 0.073 & 1.000 & 1.000 & 1.000  & 0.059 & 0.176 & 0.655 \\
		$\mathcal{H}_{1,\mathrm{C}}^{3}$ & $\boldsymbol{0.955}$ & $\boldsymbol{1.000}$ & $\boldsymbol{1.000}$ & 0.093 & 0.070 & 0.070 & 1.000 & 1.000 & 1.000 & 0.381 & 0.898 & $\boldsymbol{1.000}$ \\
		\midrule
		$\mathcal{H}_{1,\mathrm{NLQ}}^{1}$ & 0.050 & $\boldsymbol{0.174}$ & $\boldsymbol{0.305}$  & $\boldsymbol{0.080}$ & 0.072 & 0.156 & 0.102 & 0.167 & 0.626 & 0.043 & 0.082 & 0.282 \\
		$\mathcal{H}_{1,\mathrm{NLQ}}^{2}$ & $\boldsymbol{0.083}$ & $\boldsymbol{0.337}$ & 0.552 & 0.077 & 0.196 & 0.894 & 0.225 & 0.648 & 1.000 & 0.060 & 0.227 & \textbf{0.975} \\
		$\mathcal{H}_{1,\mathrm{NLQ}}^{3}$ & 0.084 & 0.420  & 0.689 & $\boldsymbol{0.250}$ & $\boldsymbol{0.983}$ & $\boldsymbol{1.000}$ & 0.502 & 0.989 & 1.000 & 0.086 & 0.532 & $\boldsymbol{1.000}$ \\
		\midrule
		$\mathcal{H}_{1,\mathrm{NLT}}^{1}$ & 0.007 & 0.039 & 0.041 & $\boldsymbol{0.074}$ & $\boldsymbol{0.083}$ & 0.047 & 0.098 & 0.145 & 0.486 & 0.039 & 0.067 & $\boldsymbol{0.190}$ \\
		$\mathcal{H}_{1,\mathrm{NLT}}^{2}$ & 0.010 & 0.044 & 0.046 & 0.067 & 0.131 & 0.767 & 0.253 & 0.655 & 1.000 & $\boldsymbol{0.069}$ & $\boldsymbol{0.244}$ & $\boldsymbol{0.961}$ \\
		$\mathcal{H}_{1,\mathrm{NLT}}^{3}$ & 0.010 & 0.042 & 0.067 & $\boldsymbol{0.385}$  & $\boldsymbol{0.998}$ & $\boldsymbol{1.000}$ & 0.625 & 0.997 & 1.000 & 0.180 & 0.758 & $\boldsymbol{1.000}$ \\
		\bottomrule
	\end{tabular}
	\caption{\small{Scenario S2. The description of Table \ref{tab:res_simple_S1} applies. \label{tab:res_simple_S2}}}
	\fi
\end{table}

\begin{table}[t!]
	\iffigstabs
	\setlength{\tabcolsep}{5pt}
	\centering
	\footnotesize
	\begin{tabular}{c|ccc|ccc|ccc|ccc}
		\addlinespace[-\aboverulesep] 
		\cmidrule[\heavyrulewidth]{2-13}
		\multicolumn{1}{c}{} & \multicolumn{3}{c}{KMSZ} & \multicolumn{3}{c}{PSS}  & \multicolumn{3}{c}{LZS} & \multicolumn{3}{c}{$\mathrm{PCvM}$} \\	
		\midrule
		$n$ & $50$ & $100$ & $250$ & $50$ & $100$ & $250$ & $50$ & $100$ & $250$ & $50$ & $100$ & $250$ \\
		\midrule 
		$\mathcal{H}_{0, \mathrm{NE}}$ & 0.006 & $\boldsymbol{0.036}$ & 0.026 & $\boldsymbol{0.046}$ & 0.071 &   $\boldsymbol{0.052}$ & $\boldsymbol{0.054}$ & $\boldsymbol{0.049}$ & $\boldsymbol{0.040}$ & $\boldsymbol{0.047}$ & $\boldsymbol{0.041}$ & $\boldsymbol{0.037}$ \\
		\midrule
		$\mathcal{H}_{1,\mathrm{FR}}^{1}$ & 0.010 & 0.041 & 0.052 & $\boldsymbol{1.000}$ & $\boldsymbol{1.000}$ & $\boldsymbol{1.000}$ & 0.058 & 0.073 & 0.107 & 0.055 & 0.108  & 0.398 \\	
		$\mathcal{H}_{1,\mathrm{FR}}^{2}$ & 0.014 & 0.040 & 0.056 & $\boldsymbol{1.000}$ & $\boldsymbol{1.000}$ & $\boldsymbol{1.000}$ & 0.063 & 0.079 & 0.134 & 0.062 & 0.119 & 0.582 \\
		$\mathcal{H}_{1,\mathrm{FR}}^{3}$ & 0.027 & 0.044 & 0.058 & $\boldsymbol{1.000}$ & $\boldsymbol{1.000}$ & $\boldsymbol{1.000}$ & 0.057 & 0.076 & 0.135 & 0.067 & 0.136 & 0.675 \\
		\midrule
		$\mathcal{H}_{1,\mathrm{C}}^{1}$ & 0.007 & 0.037 & 0.047 & 0.057 & 0.099 & 0.110 & 0.054 & 0.049 & 0.040 & $\boldsymbol{0.069}$ & $\boldsymbol{0.120}$ & $\boldsymbol{0.217}$ \\
		$\mathcal{H}_{1,\mathrm{C}}^{2}$ & 0.008 & 0.057 & 0.185 & 0.141 & 0.293 & 0.589 & 0.054 & 0.049 & 0.040 & $\boldsymbol{0.252}$ & $\boldsymbol{0.503}$ & $\boldsymbol{0.891}$ \\ 
		$\mathcal{H}_{1,\mathrm{C}}^{3}$ & 0.020 & 0.234 & 0.870 & 0.459 & 0.781 & 0.998 & 0.055 & 0.049 & 0.040 & $\boldsymbol{0.756}$ & $\boldsymbol{0.979}$ & $\boldsymbol{1.000}$ \\
		\midrule
		$\mathcal{H}_{1,\mathrm{NLQ}}^{1}$ & 0.004 & 0.027 & 0.031 & $\boldsymbol{0.061}$ & $\boldsymbol{0.132}$ & $\boldsymbol{0.374}$ & 0.083 & 0.100 & 0.261 & 0.059 & 0.080 & 0.197 \\
		$\mathcal{H}_{1,\mathrm{NLQ}}^{2}$ & 0.006 & 0.030 & 0.033 & $\boldsymbol{0.120}$ & $\boldsymbol{0.408}$ & $\boldsymbol{0.956}$ & 0.139 & 0.296 & 0.897 & 0.096 & 0.200 & 0.824 \\
		$\mathcal{H}_{1,\mathrm{NLQ}}^{3}$ & 0.007 & 0.035 & 0.036 & $\boldsymbol{0.349}$ & $\boldsymbol{0.903}$ & $\boldsymbol{1.000}$ & 0.309 & 0.785 & 1.000 & 0.201 & 0.627 & $\boldsymbol{1.000}$ \\
		\midrule
		$\mathcal{H}_{1,\mathrm{NLT}}^{1}$ & 0.005 & 0.028 & 0.034 & $\boldsymbol{0.054}$ & $\boldsymbol{0.082}$ & $\boldsymbol{0.178}$ & 0.073 & 0.082 & 0.199 & 0.052 & 0.070 & 0.156 \\
		$\mathcal{H}_{1,\mathrm{NLT}}^{2}$ & 0.005 & 0.028 & 0.033 & 0.077 & $\boldsymbol{0.252}$ & $\boldsymbol{0.986}$ & 0.131 & 0.253 & 0.940 & $\boldsymbol{0.082}$ & 0.177 & 0.816 \\
		$\mathcal{H}_{1,\mathrm{NLT}}^{3}$ & 0.008 & 0.032 & 0.030 & $\boldsymbol{0.345}$ & $\boldsymbol{0.973}$ & $\boldsymbol{1.000}$ & 0.344 & 0.873 & 1.000 & 0.207 & 0.700 & $\boldsymbol{1.000}$ \\
		\bottomrule
	\end{tabular}
	\caption{\small{Scenario S3. The description of Table \ref{tab:res_simple_S1} applies. \label{tab:res_simple_S3}}}
	\fi
\end{table}

We report some illustrative average running times of the four tests when $n=100$ and $B=1,000$. We do so only for S3, whose running times for all the tests are approximately between S1 and S2, and under $\mathcal{H}_{0,\mathrm{NE}}$ and $\mathcal{H}^3_{1,\mathrm{FR}}$ (similar results were obtained under other alternatives). For the KMSZ test (does not requires bootstrap calibration), the average running times (in seconds) were $0.0086$s ($\mathcal{H}_{0,\mathrm{NE}}$) and $0.0085$s ($\mathcal{H}^3_{1,\mathrm{FR}}$). For the PSS and LZS tests, $24.6$s and $17.5$s, and $0.5$s and $0.4$s, respectively. For the PCvM test (employs the same estimator as PSS), $0.5$s and $0.2$s. The comparison was done in a core with 1.8 GHz.\\

As a conclusion, in the considered scenarios, the PCvM test properly calibrates $\mathcal{H}_{0,\mathrm{NE}}$, is competitive against  the competing tests for all the alternatives (eventually being the most powerful in certain of them), and matches or improves the omnibus LZS and PSS tests in computational expediency.

\subsection{Composite hypothesis}
\label{sec:42}

We consider now $\mathbb{H}_1 =  L^2 \left([0,1]\right)$ and $\mathbb{H}_2 =  L^2 \left([2,3]\right)$ and
two different null (linear) hypotheses: no effects model and FLMFR. The same two nonlinear deviations from the linearity, weighted by different intensity parameters, are again considered as alternatives. Table \ref{tab:hyp_composite} summarizes all the hypothesis tested. The conclusions from the results collected in Tables \ref{tab:res_comp_S1}--\ref{tab:res_comp_S3} are the following:

\begin{itemize}
	\item As argued in Remark \ref{rem:6}, for the PCvM test based in FPCR-L1S, $\hat\lambda_{1\mathrm{SE}}$ provides better calibration of the null hypothesis than $\hat\lambda_{\mathrm{CV}}$. The latter statistic encounters serious difficulties to be calibrated, specially in S2--S3 (Tables \ref{tab:res_comp_S2}--\ref{tab:res_comp_S3}) and under $\mathcal{H}_{0,\mathrm{NE}}$. 	
	\item 
	The PCvM test based on FPCR over-rejects under irregular/sparse scenarios like S2 and S3 ($\mathcal{H}_{0,\mathrm{NE}}$ and $\mathcal{H}_{0,\mathrm{FR}}$ in Table \ref{tab:res_comp_S2}; $\mathcal{H}_{0,\mathrm{FR}}$ in Table \ref{tab:res_comp_S3}). In the case of S2, this phenomena likely arises from the overfitting (already discussed in Section \ref{sec:24}) associated with the FPCR estimator. For S3, the first scores for estimating $\beta$ are null coefficients, and therefore, the information coming from the FPC (incorrectly) suggests that $\mathcal{H}_{0,\mathrm{FR}}$ is related to a null surface (i.e., FPC suggest that $\mathcal{H}_{0,\mathrm{NE}}$ holds) and so rejection of $\mathcal{H}_{0,\mathrm{FR}}$ happens. This issue was the main motivation for developing FPCR-L1S and use it as a flexible estimator of $\beta$ within the PCvM test.
	
	\item With respect to the power, the referred over-rejection of the FPCR-based PCvM test unfairly provides greater empirical powers to this test with respect to FPCR-L1S based tests. Concerning the use of $\hat\lambda_{\mathrm{CV}}$, only marginal advantages are provided by $\hat\lambda_{\mathrm{CV}}$ in specific situations. Finally, as expected, empirical powers tends to one as $n$ and the deviation index $h$ increase. 
\end{itemize}

\begin{table}[h!]
	\iffigstabs
	\centering
	\footnotesize
	\begin{tabular}{ccl}
		\toprule
		Notation & Model & $\left(\delta_1, \delta_2, \delta_3 \right)$ \\
		\midrule
		$\mathcal{H}_{0,\mathrm{NE}}$ (no effects) & $\mathcal{Y}(t) = \mathcal{E}(t)$ & \multicolumn{1}{c}{\multirow{2}{*}{None}} \\
		\cmidrule{1-2}
		$\mathcal{H}_{0,\mathrm{FR}}$ (FLMFR) & $\mathcal{Y}(t) =   \frac{1}{2} \langle \langle \mathcal{X}, \beta\rangle \rangle + \mathcal{E}(t) $ &   \\ 
		\midrule 
		\multirow{3}{3cm}{\centering $\mathcal{H}_{1,\mathrm{NLQ}}^{h}$ (non linear, quadratic)} & \multirow{3}{6.75cm}{\centering $\mathcal{Y}(t) =  \langle \langle \mathcal{X}, \beta\rangle \rangle + \delta_h \Delta \left( \mathcal{X} \right)(t) + \mathcal{E}(t)$\newline $\Delta \left( \mathcal{X} \right)(t) =\big(\mathcal{X}^2 (a + (t - c) \frac{b - a}{d - c}) - 1 \big)$ }& S1: $\left(0.02,~0.04,~0.1\right)$   \\
		&  & S2: $\left(0.01,~0.02,~0.03\right)$   \\
		&  & S3: $\left(0.02,~0.15,~0.5\right)$  \\
		\cmidrule{1-3}
		\multirow{3}{3cm}{\centering $\mathcal{H}_{1,\mathrm{NLT}}^{h}$ (non linear, trigonometric)} &  \multirow{3}{6.75cm}{\centering $\mathcal{Y}(t) =  \langle \langle \mathcal{X}, \beta\rangle \rangle + \delta_h \Delta \left( \mathcal{X} \right)(t) + \mathcal{E}(t)$\newline $\Delta \left( \mathcal{X} \right)(t) = \left( \sin (2 \pi t)  - \cos (2 \pi t) \right) \left\| \mathcal{X} \right\|_{\mathbb{H}_1}^{2}$} & S1: $\left(0.03,~0.05,~0.1\right)$   \\
		&  & S2: $\left(0.035,~0.045,~0.055\right)$  \\
		&  & S3: $\left(0.025,~0.2,~0.45\right)$ \\
		\bottomrule
	\end{tabular}
	\caption{\small{Summary of null and alternative hypotheses, for S1--S3.}\label{tab:hyp_composite}}
	\fi
\end{table}

\begin{table}[h!]
	\iffigstabs
	\centering
	\setlength{\tabcolsep}{5pt}
	\footnotesize
	\begin{tabular}{c|ccc|ccc|ccc}
		\addlinespace[-\aboverulesep]
		\cmidrule[\heavyrulewidth]{2-10}
		\multicolumn{1}{c}{} & \multicolumn{3}{c}{FPCR} & \multicolumn{3}{c}{FPCR-L1S ($\hat\lambda_{1\mathrm{SE}}$)} & \multicolumn{3}{c}{FPCR-L1S ($\hat\lambda_{\mathrm{CV}}$)}\\
		\midrule
		$n$ & $50$ & $100$ & $250$ & $50$ & $100$ & $250$ & $50$ & $100$ & $250$ \\
		\midrule 
		$\mathcal{H}_{0,\mathrm{NE}}$ & $\mathbf{0.041}$ & $\mathbf{0.042}$ & $\mathbf{0.049}$ & $\mathbf{0.043}$ & $0.031$ & $\mathbf{0.050}$ & 0.010 & 0.014 & 0.013 \\ 
		$\mathcal{H}_{0,\mathrm{FR}}$ & $\mathbf{0.042}$ & $\mathbf{0.043}$ & $\mathbf{0.050}$ & 0.028 & $\mathbf{0.046}$ & $\mathbf{0.045}$ & 0.030 & $\mathbf{0.037}$ & $\mathbf{0.042}$ \\
		\midrule
		$\mathcal{H}_{1,\mathrm{NLQ}}^{1}$ & 0.063 & 0.091 & 0.177 & 0.045 & 0.092 & 0.179 & 0.046 & 0.080 & 0.166 \\ 
		$\mathcal{H}_{1,\mathrm{NLQ}}^{2}$ & 0.142 & 0.271 & 0.626 & 0.122 & 0.254 & 0.620 & 0.115 & 0.244 & 0.605 \\ 
		$\mathcal{H}_{1,\mathrm{NLQ}}^{3}$ & 0.596 & 0.929 & 1.000 & 0.566 & 0.917 & 1.000 & 0.568 & 0.919 & 1.000 \\ 
		\midrule
		$\mathcal{H}_{1,\mathrm{NLT}}^{1}$ & 0.048 & 0.057 & 0.115 & 0.035 & 0.059 & 0.120 & 0.037 & 0.050 & 0.106 \\
		$\mathcal{H}_{1,\mathrm{NLT}}^{2}$ & 0.087 & 0.166 & 0.642 & 0.068 & 0.155 & 0.623 & 0.068 & 0.141 & 0.608 \\ 
		$\mathcal{H}_{1,\mathrm{NLT}}^{3}$ & 0.555 & 0.953 & 1.000 & 0.496 & 0.943 & 1.000 & 0.505 & 0.941 & 1.000 \\ 
		\bottomrule	
	\end{tabular}
	\caption{\small{Scenario S1. Empirical rejection rates for the PCvM test, based on FPCR and FPCR-L1S, for $n=50, 100, 250$. Under $\mathcal{H}_{0,\mathrm{NE}}$ and $\mathcal{H}_{0,\mathrm{FR}}$, the rejection rates are boldfaced if they lie in the $95\%$-confidence interval of the nominal level, $0.05$. \label{tab:res_comp_S1}}}
	\fi
\end{table}

\begin{table}[h!]
	\iffigstabs
	\centering
	\setlength{\tabcolsep}{5pt}
	\footnotesize
	\begin{tabular}{c|ccc|ccc|ccc}
		\addlinespace[-\aboverulesep]
		\cmidrule[\heavyrulewidth]{2-10}
		\multicolumn{1}{c}{} & \multicolumn{3}{c}{FPCR} & \multicolumn{3}{c}{FPCR-L1S ($\hat\lambda_{1\mathrm{SE}}$)} & \multicolumn{3}{c}{FPCR-L1S ($\hat\lambda_{\mathrm{CV}}$)}\\
		\midrule
		$n$ & $50$ & $100$ & $250$ & $50$ & $100$ & $250$ & $50$ & $100$ & $250$ \\
		\midrule 
		$\mathcal{H}_{0,\mathrm{NE}}$ & 1.000 & 0.949 & 0.364 & 0.026 & $\mathbf{0.036}$ & $\mathbf{0.043}$ & 0.014 & 0.020 & 0.014 \\ 
		$\mathcal{H}_{0,\mathrm{FR}}$ & 0.997 & 0.876 & 0.308 & 0.091 & $\mathbf{0.047}$ & $\mathbf{0.037}$ & 0.298 & 0.109 & $\mathbf{0.047}$ \\
		\midrule
		$\mathcal{H}_{1,\mathrm{NLQ}}^{1}$ & 0.978 & 0.825 & 0.648 & 0.101 & 0.125 & 0.380 & 0.351 & 0.250 & 0.390 \\ 
		$\mathcal{H}_{1,\mathrm{NLQ}}^{2}$ & 0.993 & 0.962 & 0.990 & 0.235 & 0.463 & 0.929 & 0.504 & 0.593 & 0.934 \\ 
		$\mathcal{H}_{1,\mathrm{NLQ}}^{3}$ & 0.999 & 1.000 & 1.000 & 0.585 & 0.910 & 1.000 & 0.844 & 0.969 & 1.000 \\ 
		\midrule
		$\mathcal{H}_{1,\mathrm{NLT}}^{1}$ & 0.979 & 0.772 & 0.390 & 0.120 & 0.214 & 0.313 & 0.438 & 0.277 & 0.238 \\
		$\mathcal{H}_{1,\mathrm{NLT}}^{2}$ & 0.996 & 0.970 & 0.991 & 0.650 & 0.891 & 0.985 & 0.772 & 0.899 & 0.985 \\ 
		$\mathcal{H}_{1,\mathrm{NLT}}^{3}$ & 1.000 & 1.000 & 1.000 & 0.910 & 0.975 & 1.000 & 0.967 & 0.995 & 1.000 \\ 
		\bottomrule	
	\end{tabular}
	\caption{\small{Scenario S2. The description of Table \ref{tab:res_comp_S1} applies.\label{tab:res_comp_S2}}}
	\fi
\end{table}

\begin{table}[h!]
	\iffigstabs
	\centering
	\setlength{\tabcolsep}{5pt}
	\footnotesize
	\begin{tabular}{c|ccc|ccc|ccc}
		\addlinespace[-\aboverulesep]
		\cmidrule[\heavyrulewidth]{2-10}
		\multicolumn{1}{c}{} & \multicolumn{3}{c}{FPCR} & \multicolumn{3}{c}{FPCR-L1S ($\hat\lambda_{1\mathrm{SE}}$)} & \multicolumn{3}{c}{FPCR-L1S ($\hat\lambda_{\mathrm{CV}}$)}\\
		\midrule
		$n$ & $50$ & $100$ & $250$ & $50$ & $100$ & $250$ & $50$ & $100$ & $250$ \\
		\midrule 
		$\mathcal{H}_{0,\mathrm{NE}}$ & $\mathbf{0.042}$ & $\mathbf{0.047}$ & $\mathbf{0.046}$ & $\mathbf{0.045}$ & $\mathbf{0.043}$ & $\mathbf{0.037}$ & 0.003 & 0.008 & 0.006 \\ 	
		$\mathcal{H}_{0,\mathrm{FR}}$ & 0.082 & 0.235 & 0.957 & $\mathbf{0.044}$ & $\mathbf{0.055}$ & $\mathbf{0.057}$ & 0.009 & 0.029 & 0.125 \\
		\midrule
		$\mathcal{H}_{1,\mathrm{NLQ}}^{1}$ & 0.346 & 0.963 & 1.000 & 0.047 & 0.106 & 0.398 & 0.024 & 0.160 & 0.411 \\ 
		$\mathcal{H}_{1,\mathrm{NLQ}}^{2}$ & 0.471 & 0.988 & 1.000 & 0.054 & 0.129 & 0.597 & 0.039 & 0.203 & 0.572 \\
		$\mathcal{H}_{1,\mathrm{NLQ}}^{3}$ & 0.967 & 1.000 & 1.000 & 0.170 & 0.571 & 1.000 & 0.250 & 0.575 & 1.000 \\
		\midrule
		$\mathcal{H}_{1,\mathrm{NLT}}^{1}$ & 0.359 & 0.963 & 1.000 & 0.047 & 0.107 & 0.399 & 0.025 & 0.160 & 0.406 \\
		$\mathcal{H}_{1,\mathrm{NLT}}^{2}$ & 0.576 & 0.997 & 1.000 & 0.062 & 0.145 & 0.710 & 0.070 & 0.230 & 0.683 \\
		$\mathcal{H}_{1,\mathrm{NLT}}^{3}$ & 0.978 & 1.000 & 1.000 & 0.118 & 0.443 & 1.000 & 0.200 & 0.445 & 1.000 \\
		\bottomrule		
	\end{tabular}
	\caption{\small{Scenario S3. The description of Table \ref{tab:res_comp_S1} applies.\label{tab:res_comp_S3}}}
	\fi
\end{table}

As before, we report some illustrative average timings for S3 under the same conditions. For the PCvM-FPCR test, the timings were $0.8$s ($\mathcal{H}_{0,\mathrm{FR}}$) and $0.8$s ($\mathcal{H}_{1,\mathrm{NLQ}}$). The PCvM-FPCR-L1S ($\hat\lambda_{1\mathrm{SE}}$) test took $13.1$s and $11.9$s, and the $\hat\lambda_{\mathrm{CV}}$ variant, $11.5$s and $8.8$s. \\

As a conclusion, the obtained empirical results evidence that the PCvM test based on FPCR-L1S with $\lambda$ selected by $\hat{\lambda}_{1\mathrm{SE}}$ is a well-calibrated, flexible, and computationally efficient test that is consistent against a broad class of alternatives to the FLMFR.

\section{Real data application}
\label{sec:5}

We apply our GoF test to a real dataset with functional predictor and response (see Figure \ref{fig:real_data_1}), openly accessible as the object \texttt{aemet\_temp} from the \texttt{goffda} package. Another application is given in the SI for the dataset considered in \cite{Benatiaetal17}. Along both applications, we used $B = 10,000$ bootstrap replicates to calibrate all the bootstrap-based tests and the PCvM test was run using Algorithm \ref{algo} with FPCR-L1S, $\mathrm{EV}_p = \mathrm{EV}_q = 0.99$, and $\hat\lambda_{1\mathrm{SE}}$. For both applications, the same qualitative results were obtained with FPCR or FPCR-L2.

\begin{figure}[!h]
	\iffigstabs
	\vspace*{-0.5cm}
	\centering
	\subfloat{\includegraphics[width=.33\textwidth]{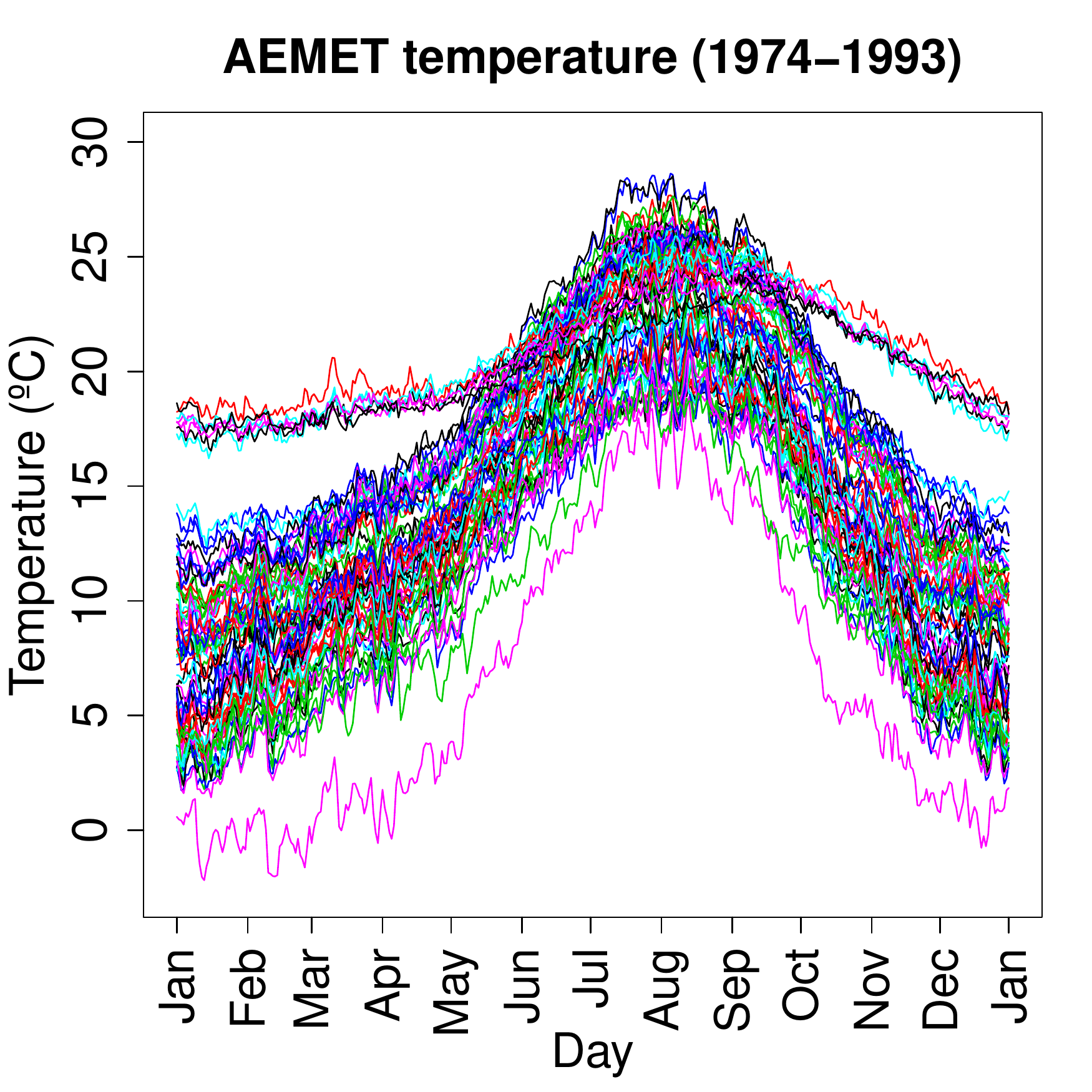}}\subfloat{\includegraphics[width=.33\textwidth]{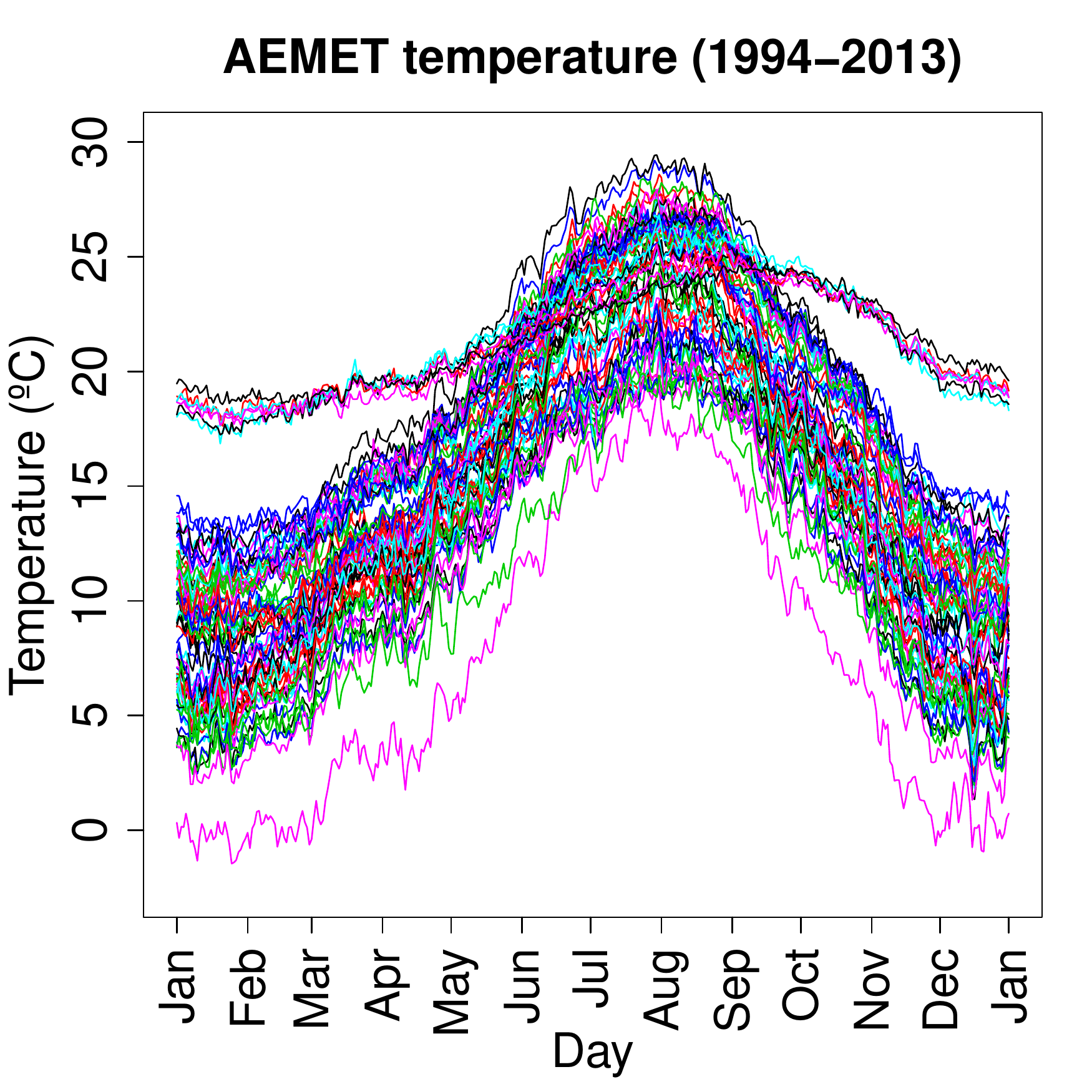}}\subfloat{\includegraphics[width=.33\textwidth]{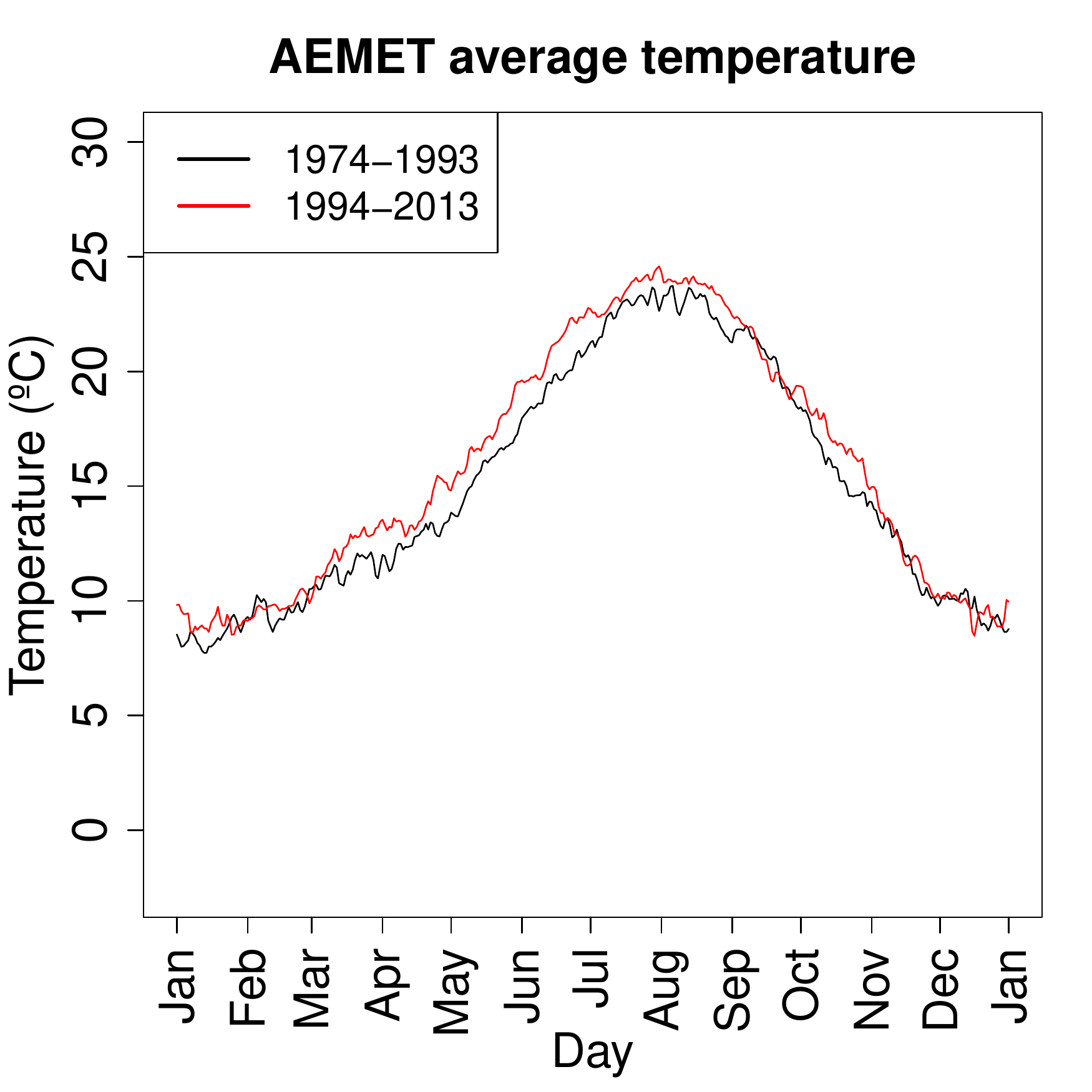}}
	\fi
	\caption{\small From left to right: samples of $\mathcal{Y}$ and $\mathcal{X}$, and sample means of $\mathcal{X}$ and $\mathcal{Y}$, indicating an increment of the average temperatures in the period 1994--2013 with respect to 1974--1983.}
	\label{fig:real_data_1}
\end{figure}

The ``AEMET temperatures dataset'' was constructed from the raw daily temperatures, along the span 1974--2013, of $n=73$ weather stations from the Meteorological State Agency of Spain (AEMET). We considered a partition of this dataset in two 20-year periods, 1974--1993 and 1994--2013, and computed the daily average temperature in each period. The aim of this partition is to explain the temperatures in the latter period ($\mathcal{Y}$) from the ones in the former ($\mathcal{X}$). Therefore, the response and predictor are valued in $\mathbb{H}_1 = \mathbb{H}_2 = L^2([0, 365])$. The functional observations are recorded in $365$ equispaced grid points in the interval $[0.5,364.5]$ and are significantly rougher than in the previous application since no presmoothing is applied. The selected stations are the same as in the \texttt{aemet} dataset of the \texttt{fda.usc} package \citep{Febrero-Bande2012} and were selected over a larger set of stations due to their consistent records and permanent locations over the 40-years\nolinebreak[4] period.\\

The PCvM test based on the data-driven $\tilde{p}=4$ and $q=3$ yielded $p\text{-value} =0.2538$ when testing the GoF of the FLMFR. Hence, the sample shows no significant evidences against the FLMFR for any sensible significance level. In addition, $\hat\beta$ in Figure \ref{fig:real_data_2} (right) reveals several interesting insights: (\textit{i}) the FLMFR mainly focuses on capturing positive correlation (positive values of $\hat\beta$; marked in red) within a $\pm90$-days band (in dashed lines) about a given time of the year, effectively corresponding to half a year; (\textit{ii}) the predominance of positive values, together with the fact that almost all records are positive, points towards a general temperature increment on the 1994--2013 span with respect to 1974--1993; (\textit{iii}) some of the visible temperature increments in the lower right panel of Figure \ref{fig:real_data_1}, such as in Apr--May and Oct--Nov, are identified with the horizontal bands spanning the same periods on the right plot of Figure \ref{fig:real_data_2}, for which there are almost no negative values of $\hat\beta$. We remark that the possible spatial dependence of the data was not taken into account in the analysis.

\begin{figure}[h!]
	\iffigstabs
	\vspace*{-0.5cm}
	\centering
	\includegraphics[width=0.5\textwidth,trim={0cm 0cm 0cm 0.5cm},clip]{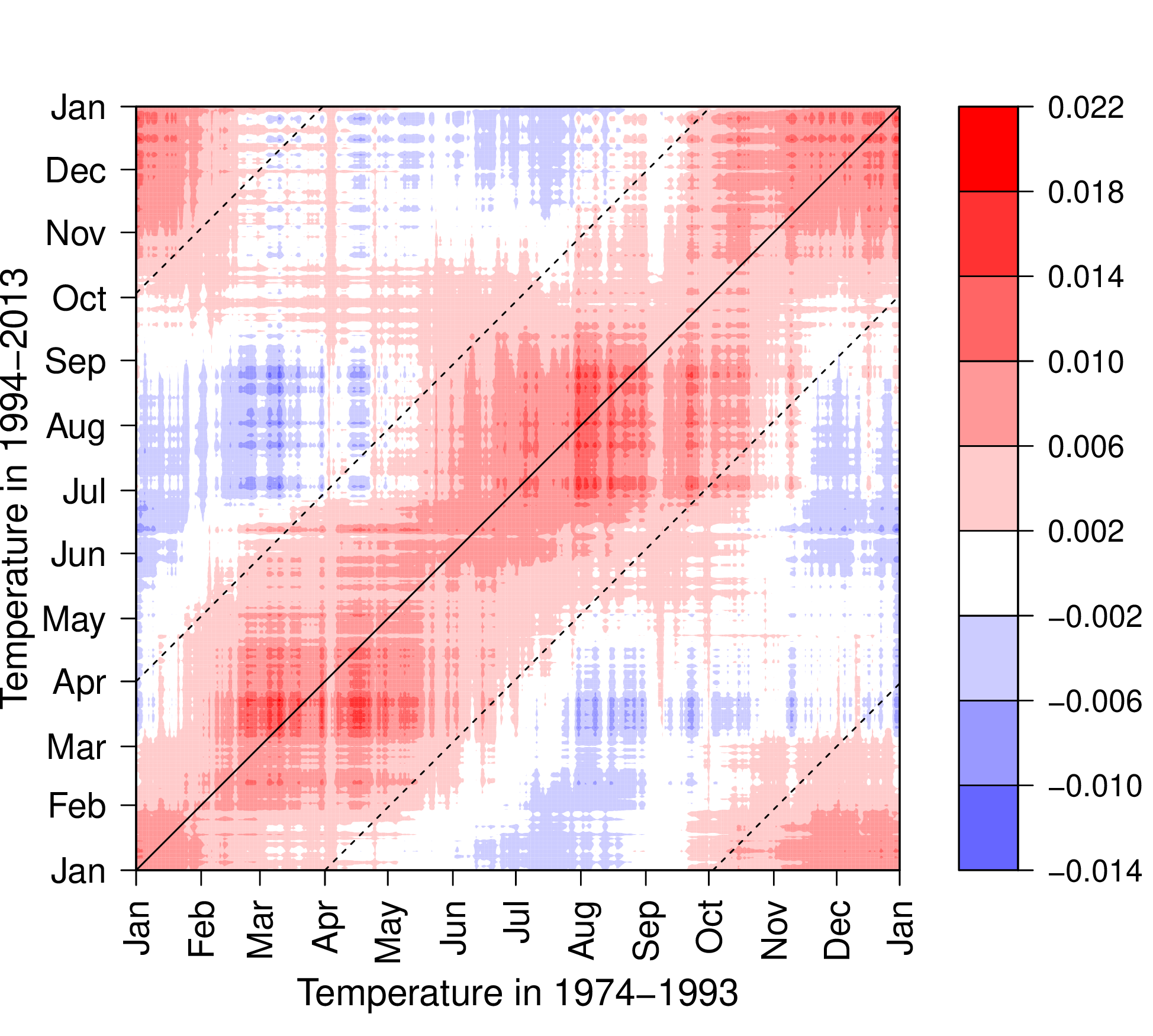}
	\fi
	\caption{\small FPCR-L1S estimator $\hat\beta$ for the AEMET temperatures dataset. Note how $\hat\beta$ reflects the smoothness of the data, inherited by the FPC.}
	\label{fig:real_data_2}
\end{figure}

One may wonder whether $\hat{\beta}$ (Figure \ref{fig:real_data_2}) is associated to a simpler FLMFR. The answer appears to be negative, as the following attempted simplifications evidence: (\textit{i}) $\mathcal{H}_0:\beta=0$ was rejected by the KMSZ, PSS, LZS, and PCvM tests with null $p$-values; (\textit{ii}) $\mathcal{H}_0:\beta=\hat{b}$, where $\hat{b}$ stands for the average value of the $\hat\beta$ surface, was rejected by the PCvM test with null $p$-value; (\textit{iii}) $\mathcal{H}_0:\beta(s,t)=\mathbbm{1}_{\{s=t\}}$, the stationary-temperature hypothesis, was rejected by the PCvM test with $p\text{-value}=10^{-4}$; (\textit{iv}) $\mathcal{H}_0:\beta(s,t)=\tilde{\hat{\beta}}(s,t)$, where $\tilde{\hat{\beta}}$ is constructed by averaging along the periodic diagonals of $\hat\beta$, was rejected by the PCvM test with null $p$-value. Interestingly, the third analysis is congruent with the outcome of the projected ANOVA \citep{CuestaFebrero10} which, when ran using \texttt{fda.usc}'s implementation with $30$ projections, rejected the equality of the mean group curves with null $p$-value. As a conclusion, this data application not only reveals that there are no evidences against the FLMFR in the studied data, but also provides yet another evidence, within a short-term and a localized region, of a significant climate change. Analogous results were obtained presmoothing with a local linear estimator featuring a cross-validated bandwidth.

\section{Conclusions}
\label{sec:6}

We have developed a GoF test for assessing the composite null hypothesis of the FLMFR. Our statistic: (\textit{i}) is based on a characterization of the null hypothesis in terms of finite-dimensional directions; (\textit{ii}) can be regarded as a weighted quadratic norm of the coefficients of the residuals in a truncated basis of $\mathbb{H}_2$; (\textit{iii}) neatly extends a previous proposal for the FLMSR. Furthermore, together with a novel estimator for the FLMFR and the use of several convenient computational procedures, we can achieve an expedient bootstrap calibration of the test statistic. Empirical results show that, in the studied scenarios, the test calibrates adequately the composite null hypothesis and detects a variety of linear and nonlinear alternatives. In addition, it is competitive against previous proposals for testing the significance of the functional predictor.\\

As noted, the PCvM statistic only depends on the functional residuals. Hence, the formulated test could be extended to alternative (possibly non-linear) regression models, provided that reliable estimators exist for them. Evident extensions are the testing of the FLMFR in the presence of several functional covariates and the testing of the functional linear model with functional response and scalar predictor.

\section*{Supporting information}

An extra real data application is provided in the supporting information.

\section*{Acknowledgements}

The authors gratefully thank Prof. Manuel Febrero-Bande for discussions at early stages of the project and for providing access to the dataset of raw AEMET temperatures. The first author acknowledges support from grants PGC2018-097284-B-I00 and IJCI-2017-32005 from the Spanish Ministry of Economy and Competitiveness (co-funded with FEDER funds). The second author acknowledges support from grant PGC2018-099549-B-I00 from the same agency. The first and fourth authors acknowledge financial support from grant MTM2016-76969-P also from the same agency. The third author acknowledges support through the Severo Ochoa Program from the Goverment of the Principality of Asturias (grant PA-20-PF-BP19-053). The authors gratefully acknowledge the computing resources of the Supercomputing Center of Galicia (CESGA). Comments by an Associate Editor and a referee are gratefully acknowledged.

\appendix

\section{Proofs}
\label{sec:poofs}

\begin{proof}[Proof of Lemma \ref{lemma:1}]
	We proceed by proving equivalences by pairs. First of all, the equivalence of \ref{lemma:1:1} and \ref{lemma:1:2} can be derived straightforwardly by the definition of $m ({\scriptstyle\mathcal{X}}) = \mathbb{E}[\mathcal{Y}| \mathcal{X}={\scriptstyle{\mathcal{X}}}]$ under $\mathcal{H}_0$. The equivalence of \ref{lemma:1:2} and \ref{lemma:1:3} follows by applying Lemma 2.1(A) in \cite{Patileaetal20} pointwisely to each point $t$ of $\mathcal{Z}(t):=(\mathcal{Y}-\langle\langle\mathcal{X},\beta\rangle\rangle)(t)$. The implication \ref{lemma:1:3} $\implies$ \ref{lemma:1:4} is trivial from the linearity of the inner product and the conditional expectation. On the other hand, the converse implication follows by taking $\gamma_{\mathcal{Y}}$ in the orthonormal basis of $\mathbb{H}_2$, $\{\Phi_k\}_{k=1}^\infty$. Then, $\mathcal{Z}(t)=\sum_{k=1}^\infty z_k\Phi_k(t)$ a.s. with $\mathbb{E}[z_k|\langle \mathcal{X}, \gamma_{\mathcal{X}} \rangle_{\mathbb{H}_1} = u] = 0$ for all $k\geq1$ and a.e. $u \in \mathbb{R}$, from where \ref{lemma:1:4} follows. Finally, the equivalence between \ref{lemma:1:4}  and \ref{lemma:1:7} arises due to the equivalence between the (real-valued and real-conditioned) conditional expectation and the integrated regression function (see, e.g., page 615 in \cite{Stute97}).
\end{proof}

\begin{proof}[Proof of Lemma \ref{lemma:2}]
	By applying pointwisely Lemma 2.1(A) in \cite{Patileaetal20}, \ref{lemma:1:3} is equivalent to \textit{iii'}, which replaces ``$\forall \gamma_{\mathcal{X}} \in \mathbb{S}_{\mathbb{H}_1}$'' by
	``$\forall \gamma_{\mathcal{X}} \in \mathbb{S}_{\mathbb{H}_1,\{\Psi_j\}_{j=1}^\infty}^{p-1}$ and for all $p\geq 1$''. As in the proof of Lemma \ref{lemma:1}, \textit{iii'} $\implies$ \textit{iv'} trivially, and the converse follows by similar arguments. The equivalence between \textit{iv'} and \textit{v'} is also analogous.
\end{proof}

\begin{proof}[Proof of Lemma \ref{lemma:3}]
	Let $\bar{\mathbf{x}}:=\mathbf{x}/\|\mathbf{x}\|\in\mathbb{S}^{q-1}$ and $\bar{\mathbf{y}}:=\mathbf{y}/\|\mathbf{y}\|\in\mathbb{S}^{q-1}$ for $\mathbf{x}\neq\mathbf{0}$ and $\mathbf{y}\neq\mathbf{0}$ (otherwise the result is trivial). Consider then the tangent-normal decomposition $\boldsymbol{\omega}=t\bar{\mathbf{x}}+(1-t^2)^{1/2}\mathbf{B}_{\bar{\mathbf{x}}}\boldsymbol{\xi}$ (as given, e.g., in Lemma 2 in \cite{Garcia-Portugues:dirlin}), where $t\in[-1,1]$, $\boldsymbol{\xi}\in\mathbb{S}^{q-2}$, and $\mathbf{B}_{\bar{\mathbf{x}}}$ is a $q\times (q-1)$ semi-orthogonal matrix such that $\mathbf{B}_{\bar{\mathbf{x}}}'\mathbf{B}_{\bar{\mathbf{x}}}=\mathbf{I}_{q-1}$ and $\mathbf{B}_{\bar{\mathbf{x}}}\mathbf{B}_{\bar{\mathbf{x}}}'=\mathbf{I}_q-\bar{\mathbf{x}}\bar{\mathbf{x}}'$. Then, the integral can be rewritten as
	\begin{align*}
	\|\mathbf{x}\|\|\mathbf{y}\|\int_{\mathbb{S}^{q-1}} (\bar{\mathbf{x}}'\boldsymbol{\omega}) (\bar{\mathbf{y}}'\boldsymbol{\omega}) \,\mathrm{d} \boldsymbol{\omega}&= \|\mathbf{x}\|\|\mathbf{y}\|\int_{\mathbb{S}^{q-2}}\int_{-1}^1 t \left(t\bar{\mathbf{y}}'\bar{\mathbf{x}}+(1-t^2)^{1/2}\mathbf{B}_{\bar{\mathbf{x}}}\boldsymbol{\xi}\right) (1-t^2)^{(q-3)/2}\,\mathrm{d} t\,\mathrm{d} \boldsymbol{\xi}\\
	&=\|\mathbf{x}\|\|\mathbf{y}\|\bar{\mathbf{x}}'\bar{\mathbf{y}} \int_{\mathbb{S}^{q-2}}\,\mathrm{d} \boldsymbol{\xi}\times\int_{-1}^1  t^2(1-t^2)^{(q-3)/2}\,\mathrm{d} t,
	\end{align*}
	where symmetry simplifies the first integral. The result follows from $\int_{\mathbb{S}^{q-2}}\,\mathrm{d}\boldsymbol{\xi}=2\pi^{(q-1)/2}/\allowbreak\Gamma \left((q-1)/2\right)$
	and $\int_{-1}^1  t^2(1-t^2)^{(q-3)/2}\,\mathrm{d} t=\sqrt{\pi} \Gamma \left((q-1)/2\right)/\allowbreak (2 \Gamma \left(q/2+1\right))$.
\end{proof}

\begin{proof}[Proof of Lemma \ref{lem:4}]
	Assume $n\geq2$, as if $n=1$ trivially $\mathbf{A}_\bullet=2\pi^{p/2}/\Gamma(p/2)>0$. From the cases described in \eqref{eq:Aijr}, and since the coefficients $\{\mathbf{x}_{i,p}\}_{i=1}^n$ are pairwise distinct, it follows that
	\begin{align*}
	\mathbf{A}_\bullet	=\frac{\pi^{p/2-1}}{\Gamma(p/2)}\left\{\sum_{r=1}^n\mathbf{A}_r+\pi\mathbf{I}_n\right\},\quad(\mathbf{A}_r)_{ij}:=\left\lbrace \begin{array} {ll} \pi, & \text{if } i = r \text{ or } j=r, \\ \eqref{spherical_wedge_eq}, & \mathrm{otherwise,} \end{array}
	\right.
	\end{align*}
	where $\mathbf{e}_r$ stands for the $r$-th canonical vector in $\mathbb{R}^n$. The matrices $\mathbf{A}_r$ have a clear block structure. For example, if $r=n$, then
	$
	\mathbf{A}_r=\left(
	\mathbf{B}_r,\: \pi\mathbf{1}_{n-1};\:
	\pi\mathbf{1}_{n-1}',\: \pi
	\right),
	$
	where $\mathbf{1}_{n-1}$ is a vector of $n-1$ ones and $(\mathbf{B}_r)_{k\ell}:=A^{(\measuredangle)}_{o_ko_\ell r}$, with
	indexes $o_i:=i + \mathbbm{1}_{\{i \leq r\}}$, $i=1,\ldots,n-1$. Analogous block expressions follow for $r<n$, yet more cumbersome since $\mathbf{B}_r$ is split into four blocks. In any case, for any $r=1,\ldots,n$, given $\mathbf{v}\in\mathbb{R}^n$, then
	$
	\mathbf{v}'\mathbf{A}_r\mathbf{v}=\mathbf{v}_{-r}'\mathbf{B}_r\mathbf{v}_{-r}+\pi\Big[v_r^2+2v_r\sum_{\substack{j=1\\ j\neq r}}^nv_j\Big]
	$
	and, as a consequence,
	$
	\mathbf{v}'\left\{\sum_{r=1}^n\mathbf{A}_r+\pi\mathbf{I}_n\right\}\mathbf{v}=\sum_{r=1}^n\mathbf{v}_{-r}'\mathbf{B}_r\mathbf{v}_{-r}+2\pi\big(\sum_{j=1}^nv_j\big)^2.
	$
	Therefore, the sum is positive for any $\mathbf{v}\neq\mathbf{0}$ if the matrices $\mathbf{B}_r$, $r=1,\ldots,n$, are positive semi-definite. \\
	
	Set $\mathbf{y}_k:=(\mathbf{x}_{k,p}-\mathbf{x}_{r,p})/\|\mathbf{x}_{k,p}-\mathbf{x}_{r,p}\|\in\mathbb{S}^{p-1}$ for $k=1,\ldots,n$, $k\neq r$, and $p\geq 1$. From \eqref{spherical_wedge_eq}, 
	$(\mathbf{B}_r)_{k\ell}=\psi\left(\cos^{-1}(\mathbf{y}_k'\mathbf{y}_\ell)\right)$, with $\psi(\theta)=\pi-\theta$, $\theta\in[0,\pi]$. Define $\tilde{\psi}(\theta):=\psi(\theta)/(2\pi)-1/4$. If $p\geq 2$, from the asymptotic distribution of the Ajne's statistic \citep[page 172]{Prentice1978},
	\begin{align*}
	& \tilde{\psi}(\theta)=\sum_{k=1}^\infty \frac{4(k-1)+p}{p-2}b_{2k-1}^2C_{2k-1}^{(p-2)/2}(\cos\theta), \  b_{2k-1}=\frac{2^{p-2} \Gamma(p/2) \Gamma(k-1+p/2)(2k-2)!}{(-1)^{k-1}\pi(k-1)! (2k+p-3)!},
	\end{align*}
	where $C_k^\alpha$ denotes the Gegenbauer polynomial of index $\alpha$ and order $k$ (when $p=2$, we use implicitly that $\lim_{\alpha\to0}C_k^\alpha(\cos\theta)/\alpha=(2/k)\cos(k\theta)$). Therefore, the Gegenbauer coefficients of $\tilde{\psi}$ are non-negative (positive if odd; null if even) and, due to the properties of the Gegenbauer polynomials, so do are the coefficients of $\psi$. Then, the characterization by \cite{Schoenberg1942} entails that $\psi$ is definite positive. This implies that, for any collection of points $\mathbf{z}_1,\ldots,\mathbf{z}_m\in\mathbb{S}^{p-1}$, for any $m\geq 2$ and $p\geq 2$, the matrix $\left(\psi\left(\cos^{-1}(\mathbf{z}_k'\mathbf{z}_\ell)\right)\right)_{k,\ell=1,\ldots,m}$ is positive semi-definite. When $p=1$, recall that $y_k\in\{-1,+1\}$ and $(\mathbf{B}_r)_{k\ell}=\pi\delta_{y_ky_\ell}$, so $\mathbf{B}_r$ can be rearranged as
	$\big(
	\pi\mathbf{1}_{n_-\times n_-},\: \mathbf{0}_{n_-\times n_+};\:
	\mathbf{0}_{n_+\times n_-},\: \pi\mathbf{1}_{n_+\times n_+}
	\big)$,
	where $n_\pm$ denotes the number of $y_k$'s equal to $\pm1$. Trivially, $(\mathbf{B}_r)_{k\ell}$ is rank $2$ with non-null eigenvalues $n_+\pi$ and $n_-\pi$. As a consequence, $\mathbf{B}_r$ is positive semi-definite for all $r=1,\ldots,n$, $n\geq2$, $p\geq1$.
\end{proof}


\fi
\ifsupplement


\newpage
\title{Supporting information for ``A goodness-of-fit test for the functional linear model with functional response''}
\setlength{\droptitle}{-1cm}
\predate{}%
\postdate{}%
\date{}

\author{Eduardo Garc\'ia-Portugu\'es$^{1,2}$, Javier \'Alvarez-Li\'ebana$^{3}$,\\
	Gonzalo \'Alvarez-P\'erez$^{4,6}$, and Wenceslao Gonz\'alez-Manteiga$^{5}$}

\footnotetext[1]{Department of Statistics, Carlos III University of Madrid (Spain).}
\footnotetext[2]{UC3M-Santander Big Data Institute, Carlos III University of Madrid (Spain).}
\footnotetext[3]{Department of Statistics and Operations Research and Mathematics Didactics, University of Oviedo (Spain).}
\footnotetext[4]{Department of Physics, University of Oviedo (Spain).}
\footnotetext[5]{Department of Statistics, Mathematical Analysis and Optimization, University of Santiago de Compostela (Spain).}
\footnotetext[6]{Corresponding author. e-mail: \href{mailto:gonzaloalvarez@uniovi.es}{gonzaloalvarez@uniovi.es}.}

\maketitle

\begin{abstract}
	This supporting information contains an extra real data application.
\end{abstract}
\begin{flushleft}
	\small\textbf{Keywords:} Bootstrap; Cram\'er--von Mises statistic; Functional data; Regularization.
\end{flushleft}


The ``Ontario dataset'', constructed by the authors of \cite{Benatiaetal17}, contains the hourly electricity consumption ($\mathcal{Y}$; measured in gigawatts) and smoothed temperature ($\mathcal{X}$; Celsius degrees) in the province of Ontario (Canada). More precisely, it features a set of $n=368$ daily curves on 2010--2014, where only summer months are taken into account, while weekends and holidays are discarded (hence, the $i$-th datum is not necessarily consecutive in time to the $(i+1)$-th). The response is valued in $\mathbb{H}_2 = L^2 \left([0,24] \right)$ and discretized in $25$ equispaced grid points. Each temperature curve is valued in $\mathbb{H}_1 = L^2 \left([-24,48] \right)$ and discretized in $73$ equispaced grid points. The interval $[-24,48]$ accounts for a $3$-days window that is considered since the past and future temperatures of a given day may influence the demand of energy on that day. Thus, the response is also regressed on 24 past and future hours. The raw temperature records are smoothed by a local polynomial regression on a weighted average of the temperatures of 41 Ontarian cities, producing the smoothed temperature, finally shifted so its minimum is set to 0$^\circ$.\\

We check whether there exists a linear relation in the Ontario dataset. This is inspired by the data application in \cite{Benatiaetal17}, where a FLMFR featuring several seasonal dummies is considered. Therefore, testing the GoF of the ``canonical'' FLMFR allows to evaluate if a seasonal-free simplified model succeeds in describing the daily electricity consumption from the temperature alone. Based on the data-driven selection of $\tilde{p} = 7$ and $q = 4$, the PCvM test gave null $p$-value, rejecting emphatically the FLMFR. When testing for significance, the KMSZ, PSS, LZS, and PCvM tests clearly rejected with null $p$-values. Hence, a nontrivial and nonlinear functional relation between daily electricity consumption and the temperature is evidenced, and the seasonal-free version of \cite{Benatiaetal17}'s model is shown to be inadequate for modeling such relation.  \\

A referee and Associate Editor pointed out that the presence of temporal dependence in the data violates the iid assumption of our GoF test. Indeed, the data construction inherited from \cite{Benatiaetal17} employs $3$-days overlapping windows that notably increases the serial dependency of the functional records. In order to investigate if this dependency was the leading rejection cause of the FLMFR, we have run our test retaining the $3$-days windows but ensuring there are no overlaps in the observations. That is, we have considered only the curves for \textit{day 1} (includes days 0, 1, 2), \textit{day 4} (days 3, 4, 5), \textit{day 7} (days 6, 7, 8), etc., properly handling weekends and holidays. The results are the same as in the original application: the FLMFR is emphatically rejected (null $p$-values) for the three possible subsettings of non-overlapping data and for different estimators. The no effect hypothesis is also rejected with null $p$-values. From this analysis, we are confident that the rejections with the original data are not primarily driven by temporal dependence (though still present in the non-overlapping data, e.g., by annual periodicity), and that a reduction in the complexity of the model in \cite{Benatiaetal17} through a seasonal-free version is not possible.

\begin{figure}[h!]
	\iffigstabs
	\vspace*{-0.5cm}
	\centering
	\includegraphics[width=\textwidth,trim={0cm 0.55cm 0cm 0.5cm},clip]{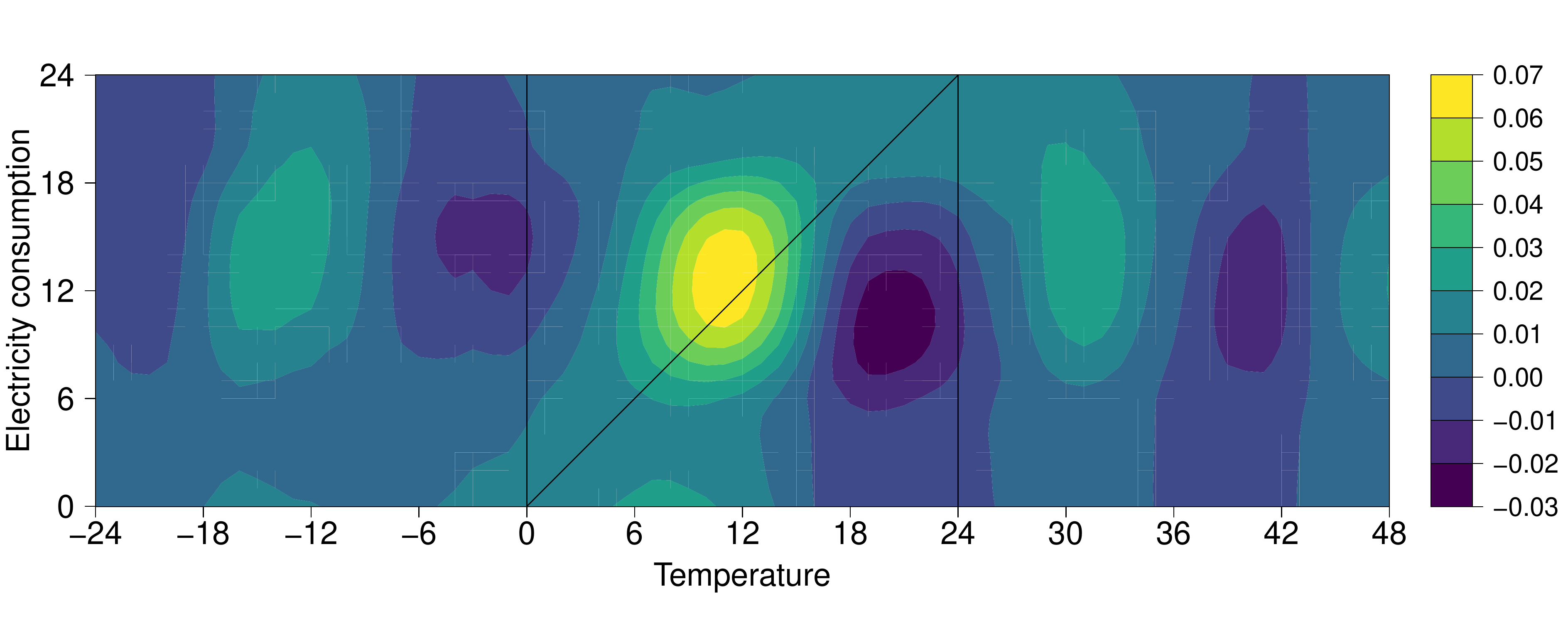}
	\fi
	\caption{\small FPCR-L1S estimator $\hat\beta$ for the Ontario dataset. Note how $\hat\beta$ reflects the smoothness of the data, inherited by the FPC. The plot is coherent with Figure 11 in \cite{Benatiaetal17}, yet ours is less centered at the diagonal, probably since no seasonal dummies were considered for fitting the FLMFR.}
	\label{fig:real_data_3}
\end{figure}

\fi

\end{document}